\documentclass[journal]{IEEEtran}

\usepackage{amsmath,amssymb,amsfonts}
\usepackage{amsthm}

\usepackage{graphicx}
\usepackage[caption=false,font=footnotesize]{subfig}
\usepackage{multicol}
\usepackage{booktabs}
\usepackage{array}

\usepackage{algorithm}
\usepackage{algpseudocode}

\usepackage{listings}
\usepackage{verbatim}
\usepackage{fancyvrb}
\usepackage{xcolor}
\usepackage[normalem]{ulem}

\usepackage{cite}
\usepackage{url}
\usepackage{hyperref}

\DeclareUnicodeCharacter{221E}{\ensuremath{\infty}}

\hypersetup{
    colorlinks=true,
    linkcolor=blue,
    citecolor=blue,
    urlcolor=blue
}

\definecolor{loginfo}{rgb}{0,0,1}
\definecolor{logwarning}{rgb}{1,0.5,0}
\definecolor{logcritical}{rgb}{1,0,0}
\definecolor{logsuccess}{rgb}{0,0.6,0}
\definecolor{logpolicy}{rgb}{0.5,0.2,0.8}
\definecolor{logcpu}{rgb}{0.8,0.6,0}
\definecolor{logmem}{rgb}{0.4,0.2,0.8}
\definecolor{logacc}{rgb}{0,0.5,0.5}
\definecolor{logresp}{rgb}{0.6,0.2,0.2}
\definecolor{loglat}{rgb}{0.7,0.7,0}

\begin{document}

\title{Green Deep Reinforcement Learning for IoT Edge Intrusion Detection}

\author{
Saeid Jamshidi,
Foutse Khomh,
Rolando Herrero,
Omar Abdul-Wahab,
and Martine Bellaiche%
\thanks{Saeid Jamshidi, Foutse Khomh, Omar Abdul-Wahab, and Martine Bellaiche are with the Department of Computer and Software Engineering, Polytechnique Montréal, 2500 Chemin de Polytechnique, Montréal, Québec H3T 1J4, Canada.}%
\thanks{Rolando Herrero is with the College of Engineering, Northeastern University, Boston, Massachusetts, USA.}%
\thanks{Corresponding author: Saeid Jamshidi (email: saeid.jamshidi@polymtl.ca).}
}

\markboth{IEEE Transactions on XXXXX, Vol. XX, No. XX, 2026}%
{Jamshidi \MakeLowercase{\textit{et al.}}: Green Deep Reinforcement Learning for IoT Edge Intrusion Detection}

\maketitle

\begin{abstract}
The rapid expansion of the Internet of Things (IoT) has intensified cybersecurity challenges, particularly in detecting and mitigating Distributed Denial-of-Service (DDoS) attacks at the network edge. Traditional Intrusion Detection Systems (IDSs) remain limited by static signatures, dependence on labeled data, poor adaptability to evolving and zero-day attacks, and high computational overhead on resource-constrained edge gateways. Moreover, most Deep Reinforcement Learning (DRL)-based IDS studies prioritize detection performance while overlooking energy consumption and carbon impact. To address these limitations, this paper proposes two carbon-aware DRL-based IDS: DeepEdgeIDS, a label-free Autoencoder-DQN architecture for anomaly-guided online mitigation, and AutoDRL-IDS, a supervised LSTM-DQN model for temporally informed detection and response. Both systems incorporate multi-objective reward functions that jointly consider security performance, response latency, energy consumption, memory utilization, and estimated carbon emissions, while using learning-paradigm-specific detection feedback. AutoDRL-IDS employs ground-truth-dependent detection metrics during supervised training, whereas DeepEdgeIDS relies on anomaly confidence and post-mitigation traffic stabilization for label-free online learning. The proposed systems are analyzed theoretically and evaluated experimentally on physical IoT edge gateways under DDoS traffic. Results show that AutoDRL-IDS achieves 94\% detection accuracy, while DeepEdgeIDS attains 98\% offline evaluation accuracy and demonstrates stronger adaptability to previously unseen attack patterns. 
\end{abstract}

\begin{IEEEkeywords}
Internet of Things, intrusion detection system, deep reinforcement learning, autoencoder, long short-term memory, DDoS detection, edge computing, sustainable cybersecurity, carbon-aware artificial intelligence.
\end{IEEEkeywords}

\maketitle

\section{Introduction}
\label{Introduction}
The Internet of Things (IoT) has become a pervasive computing paradigm, connecting billions of devices across domains such as healthcare, smart cities, transportation, and industrial automation~\cite{sadhu2022internet,taherdoost2023security}. Despite its benefits, the rapid expansion of IoT has introduced substantial cybersecurity challenges because IoT environments are decentralized, heterogeneous, dynamically configured, and often composed of devices with limited processing, memory, and energy resources~\cite{wojcicki2022internet,tran2022internet}. Among the threats targeting these environments, Distributed Denial-of-Service (DDoS) attacks remain particularly severe because they can overwhelm resource-constrained devices and edge gateways, exhaust network capacity, and disrupt critical services~\cite{shah2022blockchain,gopi2022enhanced}.
Traditional Intrusion Detection Systems (IDSs) have limited effectiveness against rapidly evolving and zero-day attacks\footnote{Zero-day attacks exploit previously unknown vulnerabilities and employ unseen attack behaviors that are not represented by existing signatures and training examples.}~\cite{zhou2023collaborative}. Signature-based IDSs rely on predefined attack patterns and therefore often fail to detect previously unseen and adversarially modified threats. Anomaly-based IDSs can identify deviations from normal behavior and are consequently more suitable for detecting novel attacks; however, they may generate excessive false positives when normal traffic patterns are highly variable. These limitations are especially problematic at IoT edge gateways, where defensive mechanisms must operate under strict computational and energy constraints while responding to attacks in real time~\cite{omolara2022internet,kumari2024towards}. Therefore, effective edge-based intrusion detection requires an adaptive mechanism that can distinguish malicious traffic, select appropriate mitigation actions, and continuously respond to changes in network behavior without imposing excessive resource overhead~\cite{heidari2023internet,moustafa2023explainable,arisdakessian2022survey}.
Deep Reinforcement Learning (DRL) provides a promising foundation for adaptive intrusion detection and mitigation because an agent can continuously observe network conditions, evaluate the consequences of defensive actions, and optimize its policy through interaction with the environment~\cite{tharewal2022intrusion,feng2024security,rizzardi2023deep}. However, deploying DRL-based IDSs on resource-constrained edge gateways introduces a fundamental trade-off among detection effectiveness, adaptability, response latency, and computational efficiency~\cite{zormati2024review}. The choice of learning paradigm further affects this trade-off. Supervised methods exploit labeled attack data and can achieve stable classification performance in known traffic conditions, but their generalization may degrade when previously unseen attacks differ substantially from the training data. Unsupervised methods do not require attack labels and can detect novel behavioral deviations, but their performance depends on the quality of normal-behavior modeling and anomaly-threshold calibration~\cite{zhao2024comparison,neuer2024unsupervised,bian2022machine,liu2023unsupervised,priyadarshi2024exploring,kharbanda2024comparative}.
A further limitation of the current literature is that most DRL-based IDS studies primarily optimize security-oriented metrics, such as detection accuracy, recall, and false-positive rate, while paying limited attention to energy consumption and the associated carbon impact of continuous learning and inference at the edge~\cite{feng2023collaborative,aljebreen2023enhancing,pashaei2022early,karthikeyan2022real}. This omission is important because an IDS may achieve high detection performance while consuming excessive CPU, memory, and energy resources, thereby reducing its practical suitability for large-scale, long-term IoT deployments. Consequently, there is a need for DRL-based IDS designs that jointly address two closely related requirements: adaptability to evolving attacks and sustainability under constrained edge resources. To address this gap, this paper proposes and comparatively evaluates two carbon-aware DRL-based IDSs for DDoS detection and mitigation at IoT edge gateways: \textit{DeepEdgeIDS} and \textit{AutoDRL-IDS}. DeepEdgeIDS combines an Autoencoder (AE) with a Deep Q-Network (DQN). The AE learns latent representations of benign traffic and produces a reconstruction-error-based anomaly score, while the DQN uses this anomaly information and network-state features to select adaptive mitigation actions. Because its online reward is constructed from deployment-available signals rather than ground-truth attack labels, DeepEdgeIDS supports label-free online adaptation. In contrast, AutoDRL-IDS combines a Long Short-Term Memory (LSTM) network with a DQN. The LSTM captures temporal dependencies in labeled traffic sequences, and the DQN learns a mitigation policy using supervised detection feedback during offline training. Thus, the two systems represent complementary learning paradigms: DeepEdgeIDS emphasizes label-free adaptation to evolving and previously unseen behavior, whereas AutoDRL-IDS emphasizes stable performance in structured scenarios for which labeled data are available. Both systems are implemented and evaluated on physical edge gateways under real-time DDoS conditions. The comparative analysis considers detection accuracy, precision, recall, false-positive behavior, response time, adaptability, CPU and memory usage, energy consumption, and estimated carbon emissions. A central contribution of the study is a carbon-aware, multi-objective reward formulation that incorporates security performance together with latency, energy overhead, memory utilization, and carbon cost. The detection-related reward component differs between the two learning paradigms. AutoDRL-IDS uses ground-truth-dependent measures, including detection rate and false-positive rate, during supervised training. DeepEdgeIDS instead uses label-free signals based on anomaly confidence, post-mitigation traffic stabilization, and a blocking-cost penalty. This distinction preserves DeepEdgeIDS's label-free deployment objective while enabling both systems to optimize a shared set of sustainability-related criteria. The proposed framework also models the relationships among power consumption, execution time, energy overhead, and carbon emissions, allowing sustainability indicators to continuously influence policy learning rather than being reported only after deployment. Under standard bounded-reward and stochastic-approximation assumptions, the convergence behavior of the DQN-based learning process is analyzed using the Bellman contraction property~\cite{kadurha2025bellman}. In addition, Analysis of Variance (ANOVA)~\cite{st1989analysis} is used to examine whether the observed differences between the two systems are statistically significant across operational metrics, including detection probability, response time, latency, CPU usage, and energy consumption. The experimental findings indicate that DeepEdgeIDS achieves higher detection accuracy, faster response times, and greater adaptability to unseen attack patterns, although its continuous online adaptation incurs higher aggregate CPU and energy overhead. AutoDRL-IDS provides more stable behavior in structured attack scenarios and can be more energy-efficient under lower traffic loads. The main contributions of this work are summarized as follows:

\begin{itemize}

\item \textbf{Two carbon-aware DRL-based IDS:}
We introduce DeepEdgeIDS, a label-free AE-DQN for anomaly-guided adaptive mitigation, and AutoDRL-IDS, a supervised LSTM-DQN for temporally informed intrusion detection and response at resource-constrained IoT edge gateways.

\item \textbf{Unified sustainability-aware optimization and theoretical analysis:}
We develop multi-objective reward formulations that jointly optimize detection performance, latency, energy consumption, memory utilization, carbon emissions, and mitigation cost. The learning process is further analyzed under bounded-reward, Bellman-contraction, and stochastic-approximation assumptions.

\item \textbf{Real-world deployment and comparative evaluation:}
Both systems are implemented on physical edge gateways and evaluated under live DDoS attacks. Their performance is systematically compared across detection accuracy, zero-day adaptability, false-positive behavior, response time, computational overhead, energy efficiency, and estimated carbon emissions.

\end{itemize}

The remainder of this paper is organized as follows. Section~\ref{Related Work} reviews related IDS research. Sections ~\ref{Proposed Solutions} describe the proposed solutions. Section~\ref{Experimental setup} outlines the experimental setup and dataset. Section~\ref{EXPERIMENTS AND EVALUATION} presents evaluation results, while Section~\ref{Discussion} discusses key understanding. Section~\ref{Limitations and Future Work} outlines limitations and future work, and Section~\ref{Conclusion} concludes the paper.

\section{Related Work}
\label{Related Work}
This section provides an overview of existing IDS solutions in IoT networks, highlighting advancements in ML and DRL techniques for real-time threat detection and mitigation.
\subsection{ML-Based IDS for DDoS Detection}
Several studies have proposed intelligent ML-based frameworks to enhance DDoS detection in IoT networks. Yousuf et al. \cite{yousuf2022ddos} introduced DALCNN, an RNN-based model integrated into an SDN framework, achieving 99.98\% accuracy with enhanced detection capabilities and minimal false positives. Pandey et al. \cite{pandey2023performance} explored entropy-based methods, demonstrating that \(\phi\)-entropy with quartile-based thresholds improves detection precision but requires further enhancement in recall. Ahmad et al. \cite{ahmad2023big} proposed a big data-driven ensemble model that combines SVM, CNN, and GRU and is optimized using the Slime Mould Algorithm (SMA), achieving 98.45\% accuracy while reducing false positives and improving detection speed. Nguyen et al. \cite{nguyen2023robust} developed a hybrid learning model combining SOCNN, LOF, and iNNE, demonstrating superior detection of unknown and adversarial attacks across benchmark datasets. Gupta et al. \cite{gupta2023deep} introduced a lightweight CNN-based framework for resource-constrained IoT devices, achieving 99.38\%, accuracy and outperforming conventional ML classifiers. These works collectively highlight the effectiveness of ML-based IDS in addressing evolving DDoS threats while underscoring the need for further innovation to reduce computational overhead and enhance adaptive capabilities.
\subsection{DRL-Based IDS for DDoS Detection}
DDoS attacks in IoT remain challenging due to the limited computational resources of edge devices and the dynamic nature of threats. DRL has emerged as a promising paradigm, enabling IDS to learn and adapt to cyber threats autonomously. Feng et al. \cite{feng2023collaborative} introduced a Soft Actor-Critic (SAC)-based DRL approach that utilizes lightweight unsupervised classifiers at edge gateways to detect volumetric and stealthy low-rate DDoS attacks, achieving 91\% accuracy while maintaining low resource overhead. Liu et al. \cite{liu2022ieee} proposed a hierarchical DRL framework that mitigates multi-layer DDoS attacks, achieving 97\%, detection accuracy by leveraging cross-layer threat intelligence. Vadigi et al. \cite{vadigi2023federated} integrated federated learning with DRL to enhance scalability and privacy in IoT security, achieving 96.7\% accuracy on the NSL-KDD dataset and 99.66\% on the ISOT-CID dataset. Their dynamic attention mechanism enables real-time adaptation to evolving attack patterns without sharing raw data, preserving data privacy in decentralized environments.\\
Beyond DDoS detection, RL has been effectively employed in broader IDS scenarios. Ramana et al. \cite{ramana2022ambient} implemented a DQN-based IDS capable of detecting both binary and multi-class attacks in IoT networks, demonstrating high scalability and accuracy across the UNSW-NB-15 and CICIDS2017 datasets. Ma et al. \cite{ma2023decision} proposed a Minimax-DQN model that leverages game-theoretic principles to optimize intrusion response strategies in fog computing environments, thereby improving response time and the efficiency of attack mitigation. Zhang et al. \cite{zhang2021anti} introduced a zero-sum game-theoretic DRL model to secure IoT edge devices against adversarial behaviors, ensuring robust threat detection with minimal computational footprint.\\
DRL techniques have also demonstrated effectiveness in malware and botnet detection, which is crucial for securing IoT networks. Al-Garadi et al. \cite{al2023malbot} developed a DRL-driven botnet detection framework that achieves over 99\%, detection accuracy across different lifecycle stages of malware, addressing challenges such as model drift and evolving attack patterns. Their work underscores the potential of DRL in proactive and adaptive IoT security solutions.\\

The literature on IoT IDS has demonstrated substantial progress through ML- and DRL-based approaches, particularly in improving DDoS detection accuracy and adaptive response. However, persistent challenges remain in identifying zero-day attacks, limiting computational and energy overhead, and supporting real-time operation on resource-constrained edge gateways. To address these limitations, this work introduces two complementary DRL-based IDS: DeepEdgeIDS, a label-free AE-DQN hybrid for anomaly-guided online mitigation, and AutoDRL-IDS, a supervised LSTM-DQN model for temporally informed detection and response. Both systems are implemented and evaluated on physical edge gateways under realistic DDoS conditions. Unlike prior studies that primarily optimize security performance, the proposed systems incorporate carbon-aware, multi-objective reward functions that jointly account for detection effectiveness, response latency, resource utilization, energy consumption, and estimated carbon emissions. The resulting comparison provides new insights into the trade-offs between supervised and label-free DRL paradigms and supports the development of scalable, adaptive, and sustainable intrusion detection at the IoT edge.

\section{Proposed Solutions}
\label{Proposed Solutions}
This section introduces two proposed DRL-based IDSs, namely DeepEdgeIDS and AutoDRL-IDS.

\subsection{DeepEdgeIDS}
\label{sec:DeepEdgeIDS}
DeepEdgeIDS is a DRL-based IDS designed to detect and mitigate DDoS attacks at the edge gateways. The DRL-based IDS integrates an AE-based unsupervised feature extractor with a Deep Q-Network (DQN) for adaptive mitigation. Moreover, by combining self-learned representations with reinforcement-driven control, DeepEdgeIDS enables real-time adaptability, energy efficiency, and carbon-aware sustainability in dynamic IoT networks.
\begin{figure*}[h!]
    \centering
    \includegraphics[width=0.8\textwidth, height=0.70\textheight, keepaspectratio]{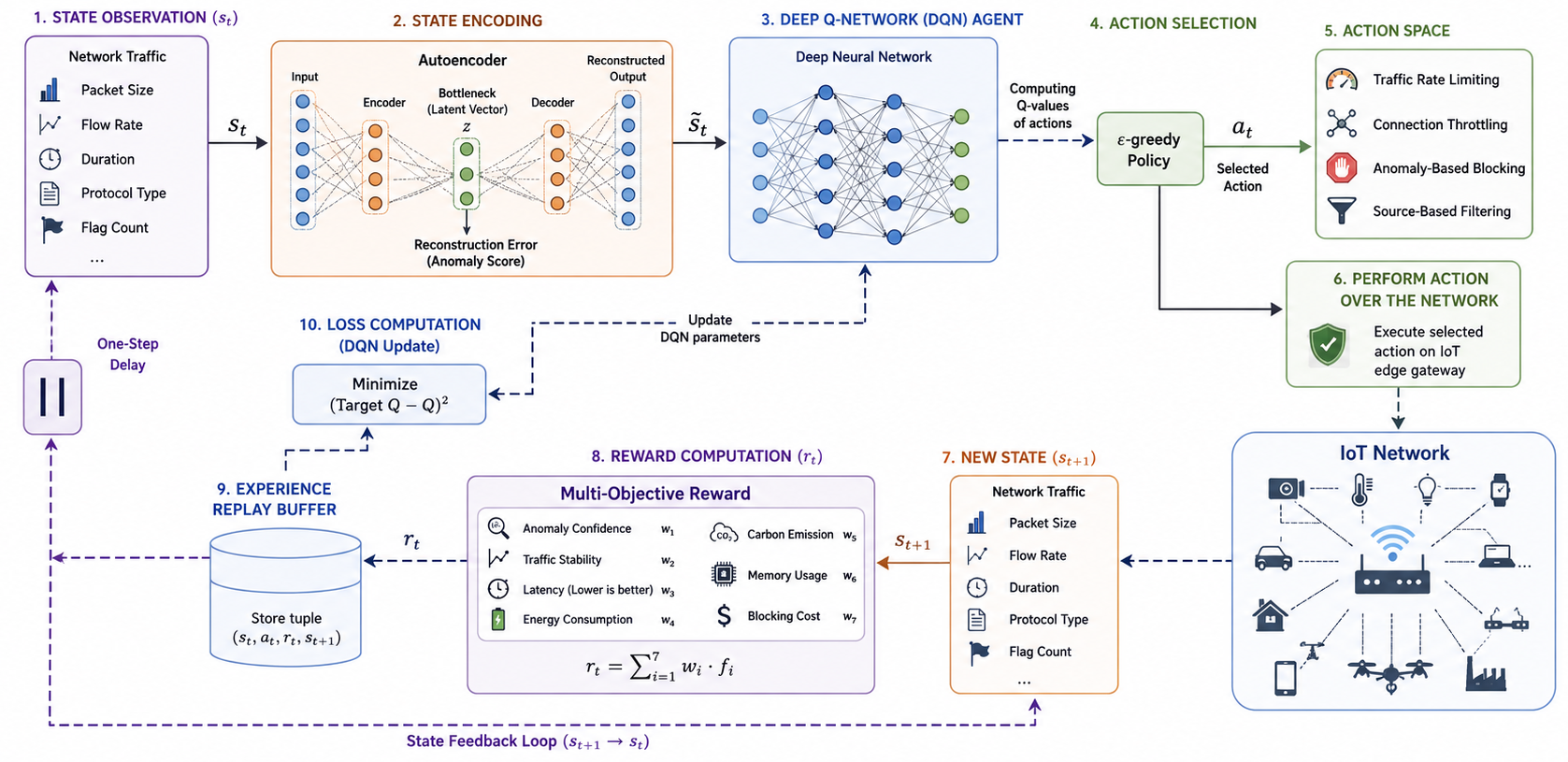}
    \caption{Overview of the proposed DeepEdgeIDS architecture for DDoS detection. }
    \label{fig:DeepEdgeIDS}
\end{figure*}
As illustrated in the Figure.~\ref{fig:DeepEdgeIDS}, DeepEdgeIDS combines AE-based feature extraction with DRL-based decision making. Network traffic is continuously monitored and encoded into latent features, with reconstruction error used to detect anomalies. Deviations are forwarded to the DQN agent, along with raw traffic statistics, which then assesses the severity and selects the optimal mitigation actions. Experience replay ensures stable learning, while a sustainability-driven reward balances detection, latency, energy, and carbon cost.
\subsubsection{AE-Based Feature Extraction}
The AE learns a compressed representation of normal traffic patterns and identifies anomalies by reconstructing the input with error. Given input features \(x_t\):
\begin{equation}
h_t = f_{\theta}^{encoder}(x_t), \quad \hat{x}_t = f_{\theta}^{decoder}(h_t),
\end{equation}
The reconstruction error is:
\begin{equation}
A_s = \|x_t - \hat{x}_t\|^2,
\end{equation}
where large \(A_s\) indicates abnormal activity. Traffic is flagged if
\begin{equation}
A_s > \tau_A \Rightarrow \text{DDoS suspected.}
\end{equation}
These AE-derived features augment the DQN state for adaptive learning.
\subsubsection{Mitigation Strategy and State Modeling}
The detection–mitigation process is modeled as a Markov decision process with state:
\begin{equation}
s_t = \{P_{rate}, SYN_{count}, ACK_{count}, A_s, h_t\},
\end{equation}
and action space:
\begin{equation}
A = \{a_1,a_2,a_3,a_4\}.
\end{equation}
Q-values are updated using:
\begin{equation}
Q(s_t,a_t) = R(s_t,a_t) + \gamma \max_{a'} Q(s_{t+1},a'),
\end{equation}
ensuring long-term optimization. 
To formalize this in control-theoretic form, the mitigation dynamics can be expressed as:
\begin{equation}
\dot{x}(t) = f(x(t)) + g(x(t))u(t),
\end{equation}
where \(x(t)\) is the network state vector, and \(u(t)\in A\) is the mitigation control input. The optimal policy $\pi^\ast$ minimizing cumulative cost:
\begin{equation}
J(\pi) = \int_0^{T} \big(\mathcal{L}(x(t),u(t)) + \zeta C_{emission}(t)\big) dt,
\end{equation}
subject to energy constraints \(E_{overhead}(t)\leq E_{max}\), satisfies the Hamilton–Jacobi–Bellman (HJB) \cite{sharpless2025dual} equation:
\begin{equation}
0 = \min_{u\in A}\left\{ \mathcal{L}(x,u) + \nabla_x V(x)^\top (f(x)+g(x)u)\right\},
\end{equation}
where \(V(x)\) is the value function approximated by the DQN, \(V(x)\approx Q(s,a)\). 

\subsubsection{Sustainability-Aware Reward Function}
\label{R_1}
Since DeepEdgeIDS operates in a label-free online setting, its reward function is formulated using deployment-available feedback rather than ground-truth-dependent metrics such as detection rate and false-positive rate. Each mitigation action is evaluated through a multi-objective reward that combines anomaly confidence, post-mitigation traffic stabilization, response latency, resource utilization, and environmental cost:
\begin{align}
R(s_t,a_t) &= 
\alpha A^{conf}_t 
+ \mu S^{stab}_t
- \lambda_L L_{resp,t} \nonumber \\
&\quad
- \delta E_{overhead,t}
- \epsilon M_{util,t}
- \zeta C_{emission,t}
- \rho B_t ,
\end{align}
where \(A^{conf}_t\) denotes the anomaly confidence derived from the AE reconstruction error, \(S^{stab}_t\) represents post-mitigation traffic stabilization, \(L_{resp,t}\) is the response latency, \(E_{overhead,t}\) is the energy overhead, \(M_{util,t}\) is the memory utilization, \(C_{emission,t}\) is the estimated carbon emission, and \(B_t\) is a blocking-cost penalty that discourages overly aggressive mitigation actions. The anomaly confidence is computed from the reconstruction error as:
\begin{equation}
A^{conf}_t =
\sigma\left(\frac{A_s(t)-\tau_A}{\tau_A+\varepsilon}\right),
\end{equation}
where \(A_s(t)\) is the AE reconstruction error, \(\tau_A\) is the anomaly threshold, \(\varepsilon\) is a small constant used to avoid division by zero, and \(\sigma(\cdot)\) is a bounded normalization function. The traffic stabilization term is defined as:
\begin{equation}
S^{stab}_t =
\frac{\max(0, P_{rate,t}-P_{rate,t+1})}{P_{rate,t}+\varepsilon},
\end{equation}
where \(P_{rate,t}\) and \(P_{rate,t+1}\) denote the packet rate before and after the selected mitigation action, respectively. A higher value of \(S^{stab}_t\) indicates that the selected mitigation action successfully reduced abnormal traffic intensity. In contrast to supervised reward formulations, DeepEdgeIDS's online reward does not require ground-truth labels. Therefore, metrics such as detection rate, false-positive rate, accuracy, precision, recall, and F1-score are used only for offline evaluation using labeled test traces, not for real-time policy learning. To ensure stability under constrained energy budgets, a Lyapunov function is introduced:
\begin{equation}
V_t = \frac{1}{2}\|Q_t - Q^\ast\|^2.
\end{equation}
Differentiating along the training trajectory yields:
\begin{equation}
\dot{V}_t = (Q_t - Q^\ast)^\top(\dot{Q}_t - \dot{Q}^\ast)
\le -\eta \|Q_t - Q^\ast\|^2,
\end{equation}
which ensures exponential convergence to the optimal policy when \(\eta>0\), assuming bounded rewards and stable stochastic approximation updates.

\subsubsection{Modeling Energy, Memory, and Carbon Cost}
Sustainability terms are modeled as:
\begin{align}
E_{overhead}(t) &= \int_{0}^{\Delta t} P(\tau) \, d\tau = P_t \cdot \Delta t, \\
M_{util}(t) &= \frac{M_{active}(t)}{M_{total}}, \\
C_{emission}(t) &= \int_0^{\Delta t} \kappa(\tau) P(\tau) \, d\tau \approx E_{overhead}(t)\kappa_t,
\end{align}
and boundedness is guaranteed since:
\begin{equation}
E_{overhead}(t) \le P_{max}\Delta t, \quad C_{emission}(t)\le \kappa_{max}E_{overhead}(t).
\end{equation}
The optimal steady-state sustainability point occurs when:
\begin{equation}
\frac{\partial R}{\partial E_{overhead}} + \zeta\kappa_t = 0 \Rightarrow E_{opt} = -\frac{\zeta\kappa_t}{\partial R/\partial E_{overhead}}.
\end{equation}
\subsubsection{Experience Replay and Convergence}
The Bellman \cite{vasfi2025channel} operator $\mathcal{T}$ satisfies:
\begin{equation}
\|\mathcal{T}Q_1 - \mathcal{T}Q_2\|_\infty \le \gamma \|Q_1 - Q_2\|_\infty,
\end{equation}
guaranteeing convergence to $Q^\ast$. Using stochastic approximation, the update:
\begin{equation}
Q_{k+1}(s,a) = (1-\eta_k)Q_k(s,a) + \eta_k\left[R_t + \gamma \max_{a'}Q_k(s',a')\right],
\end{equation}
converges a.s. if $\sum_k \eta_k = \infty$, $\sum_k \eta_k^2 < \infty$.
\subsubsection{Control-Theoretic Stability Analysis}
Let the overall closed-loop dynamics be:
\begin{equation}
x_{t+1} = f(x_t) + g(x_t)\pi_\theta(s_t),
\end{equation}
and define a Lyapunov \cite{bi2021lyapunov} function \(V(x_t) = x_t^\top Px_t\) with \(P=P^\top>0\).  
Stability holds if:
\begin{equation}
\Delta V = V(x_{t+1}) - V(x_t) \le -x_t^\top Qx_t - \pi_\theta(s_t)^\top R \pi_\theta(s_t),
\end{equation}
for some positive definite matrices \(Q,R\).  
Therefore, the learned policy $\pi_\theta$ is asymptotically stable under bounded power and carbon constraints.
\subsubsection{Adaptive Source Filtering}
Persistent threats are identified using:
\begin{equation}
P_{attack}(i)=\frac{1}{T}\sum_{t=1}^{T}\mathbb{1}(A_s(i,t)>\tau_A),
\end{equation}
and IPs with $P_{attack}(i)>\tau_P$ are blacklisted for long-term prevention.

\subsubsection{DeepEdgeIDS Pipeline}
Algorithm \ref {alg:DeepEdgeIDS} presents the closed-loop control of DeepEdgeIDS, integrating AE-based anomaly extraction with a DQN for adaptive mitigation. 
At each timestep $t$, network traffic $x_t$ is encoded as $h_t=f_{\theta}^{encoder}(x_t)$, and the reconstruction error $a_s=\|x_t-\hat{x}_t\|^2$ quantifies anomaly deviation. 
If $a_s>\tau_A$, the state $s_t=\{P_{rate},SYN_{count},ACK_{count},a_s,h_t\}$ is formed and the optimal mitigation $a_t=\arg\max_a Q(s_t,a)$ is chosen. 
Each action is mapped to a predefined policy, including rate limiting, throttling, blocking, and source filtering.  The reward is multi-objective and label-free:
\[
r_t=
\alpha A^{conf}_t
+\mu S^{stab}_t
-\lambda_L L_t
-\delta E_t
-\epsilon M_t
-\zeta C_t
-\rho B_t,
\]
balancing anomaly confidence, post-mitigation traffic stabilization, latency, energy, memory utilization, carbon cost, and blocking aggressiveness. This formulation preserves DeepEdgeIDS's label-free online operation by not requiring ground-truth labels at deployment time. Q-values are iteratively updated as 
\[
Q_{new}(s_t,a_t)=(1-\eta)Q(s_t,a_t)+\eta[r_t+\gamma\max_{a'}Q(s_{t+1},a')],
\]
with gradients $\nabla_{\theta_Q}(Q(s_t,a_t)-y_t)^2$ driving convergence toward $Q^\ast$. 
The Bellman operator $\mathcal{T}$ satisfies $\|\mathcal{T}Q_1-\mathcal{T}Q_2\|_\infty\le\gamma\|Q_1-Q_2\|_\infty$, 
and stability follows from the Lyapunov condition $\dot{V}(Q_t)\le -\eta\|Q_t-Q^\ast\|^2$. 
Energy and carbon costs are modeled as 
$E_t=P_t\Delta t$, $C_t=\int_0^{\Delta t}\kappa(\tau)P(\tau)d\tau$, 
ensuring reward-optimal energy-aware control. 
The system alternates between active mitigation ($a_s>\tau_A$) and passive monitoring ($a_s\le\tau_A$), 
enabling real-time DDoS detection with provable convergence and sustainability-aware optimization at the edge.
\begin{algorithm}[t]
\small
\caption{DeepEdgeIDS pipeline}
\label{alg:DeepEdgeIDS}
\begin{algorithmic}[1]
\State \textbf{Initialize:} AE, DQN, Replay Buffer $\mathcal{D}$
\State Set hyperparameters $\gamma, \alpha, \beta, \delta, \epsilon, \eta$
\While{Network is active}
    \State Capture traffic $x_t$, extract features $f_t$
    \State Encode: $h_t \gets f_{\theta}^{encoder}(f_t)$
    \State Compute error: $a_s = \|x_t - \hat{x}_t\|^2$
    \If{$a_s > \tau_A$} \Comment{Anomaly detected}
        \State $s_t = \{P_{rate}, SYN_{count}, ACK_{count}, a_s, h_t\}$
        \State $a_t = \arg\max_a Q(s_t,a)$
        \If{$a_t = a_1$} Limit $P_{rate}$ \Comment{Rate limiting}
        \ElsIf{$a_t = a_2$} Restrict SYN requests \Comment{Throttling}
        \ElsIf{$a_t = a_3$} Drop anomalous packets \Comment{Blocking}
        \ElsIf{$a_t = a_4$} Blacklist IPs \Comment{Source filtering}
        \EndIf
        \State Compute label-free reward $r_t$ (anomaly confidence, stabilization, latency, energy, memory, carbon)
        \State Store $(s_t,a_t,r_t,s_{t+1})$ in $\mathcal{D}$
        \If{Buffer full}
            \State Sample minibatch $(s,a,r,s')$ from $\mathcal{D}$
            \State Update $Q(s,a)$ via Bellman equation
            \State Gradient update on DQN
            \State Decay $\epsilon$ to refine policy
        \EndIf
    \Else
        \State Continue passive monitoring
    \EndIf
\EndWhile
\end{algorithmic}
\end{algorithm}

\subsection{AutoDRL-IDS}
\label{sec:AutoDRL}
\begin{figure*}[h!]
    \centering
    \includegraphics[
        width=0.80\textwidth,
        height=0.72\textheight,
        keepaspectratio
    ]{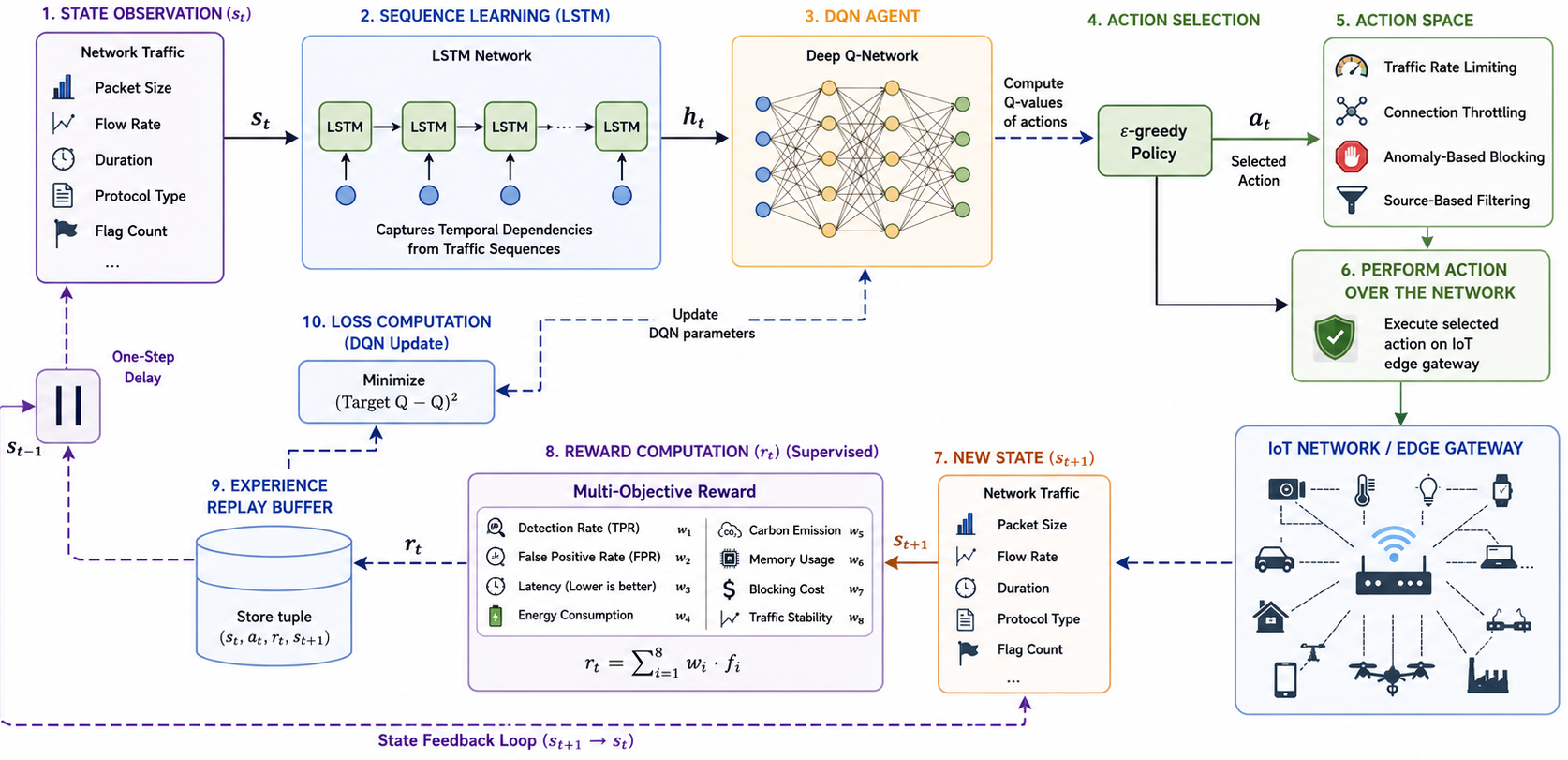}
    \caption{Overview of the proposed AutoDRL-IDS architecture, integrating LSTM-based temporal feature learning with a DQN agent for supervised, carbon-aware DDoS detection and adaptive mitigation at the IoT edge gateway.}
    \label{fig:AutoDRL-IDS}
\end{figure*}

AutoDRL-IDS is a supervised DRL-based IDS that augments the DeepEdgeIDS architecture with temporal awareness and labeled-data supervision. As illustrated in Figure~\ref{fig:AutoDRL-IDS}, it employs an LSTM network to capture sequential dependencies in network traffic and integrates the resulting temporal representation with a DQN agent for adaptive mitigation. The agent evaluates the current traffic state, selects an appropriate response from the available mitigation actions, and updates its policy using experiences stored in the replay buffer. In contrast to the label-free online operation of DeepEdgeIDS, AutoDRL-IDS bridges offline supervised training and online policy adaptation, enabling carbon-aware and sustainable real-time DDoS mitigation at the edge.
For AutoDRL-IDS, where labeled data are available during offline supervised training, the optimization objective can incorporate ground-truth-dependent detection feedback. The supervised carbon-aware optimization objective is formulated as a constrained stochastic control problem:
\begin{align}
\max_{\theta_Q,\theta_L} \quad & 
\mathbb{E}_{\pi_\theta}\!\left[
\sum_{t=0}^{T}\gamma^t
\left(
DR_t
- \lambda FPR_t
- \zeta C_{emission}(t)
\right)
\right], \nonumber\\
\text{s.t.}\quad &
E_{overhead}(t) \le E_{max}, \;
M_{util}(t) \le M_{max}.
\end{align}
Here, \(DR_t\) and \(FPR_t\) denote detection rate and false-positive rate, respectively, and are used only because AutoDRL-IDS is trained in a supervised setting with labeled data. In contrast, DeepEdgeIDS does not use \(DR_t\) and \(FPR_t\) as components of the online reward. Its online objective replaces these label-dependent terms with label-free feedback signals, namely anomaly confidence \(A^{conf}_t\) and post-mitigation traffic stabilization \(S^{stab}_t\), preserving its unsupervised deployment setting.
To handle the energy and memory constraints, we construct the Lagrangian\cite{wang2024real}:
\begin{equation}
\mathcal{L}(\theta, \lambda_E, \lambda_M) =
J(\theta)
- \lambda_E(E_{overhead}-E_{max})
- \lambda_M(M_{util}-M_{max}),
\end{equation}
where \(J(\theta)\) denotes the expected carbon-aware return, and \(\lambda_E\) and \(\lambda_M\) are non-negative Lagrange multipliers associated with the energy and memory constraints, respectively.
The corresponding gradient descent step becomes:
\begin{equation}
\nabla_\theta \mathcal{L} =
\nabla_\theta J(\theta)
- \lambda_E \nabla_\theta E_{overhead}
- \lambda_M \nabla_\theta M_{util}.
\end{equation}
This induces a carbon-weighted term that modifies the DRL update rule:
\begin{equation}
\begin{split}
Q_{t+1}(s_t,a_t) &= (1-\eta)Q_t(s_t,a_t) \\
&\quad + \eta \Big[
r_t
+ \gamma \max_{a'} Q_t(s_{t+1},a')
- \xi\nabla_\theta C_{emission}(t)
\Big],
\end{split}
\end{equation}
where \(\xi\) dynamically scales the environmental penalty, ensuring that optimization remains energy- and carbon-efficient. The equilibrium between detection performance and sustainability can be expressed through the Karush-Kuhn-Tucker \cite{liu2021policy} conditions:
\begin{align}
\nabla_\theta J(\theta^\ast)
- \lambda_E^\ast \nabla_\theta E_{overhead}
- \lambda_M^\ast \nabla_\theta M_{util} &= 0, \\
\lambda_E^\ast (E_{overhead}-E_{max}) &= 0, \\
\lambda_M^\ast (M_{util}-M_{max}) &= 0,
\end{align}
which characterize Pareto-optimal \cite{dou2024towards} learning under coupled constraints on detection, energy, memory, and carbon.

\subsection{Dynamic Reward Function and Control-Theoretic Formulation}
\label{R_2}
At each time step $t$, the carbon-augmented reward is defined based on the learning setting of each IDS. For AutoDRL-IDS, where labeled data are available during supervised training, the reward can include ground-truth-dependent detection feedback:
\begin{align}
R^{sup}_t &= \alpha DR_t - \beta FPR_t - \lambda_L L_t 
- \delta E_t - \epsilon M_t - \zeta C_t, \\
C_t &= \int_0^{\Delta t} \kappa(\tau)P(\tau)d\tau.
\end{align}
For DeepEdgeIDS, which operates in a label-free online setting, the reward replaces \(DR_t\) and \(FPR_t\) with deployment-available feedback signals:
\begin{align}
R^{unsup}_t &=
\alpha A^{conf}_t
+\mu S^{stab}_t
-\lambda_L L_t
-\delta E_t
-\epsilon M_t
-\zeta C_t
-\rho B_t, \\
C_t &= \int_0^{\Delta t} \kappa(\tau)P(\tau)d\tau.
\end{align}
Thus, both IDS models share the same sustainability penalties for latency, energy, memory, and carbon emissions, but differ in the detection-feedback component of the reward. AutoDRL-IDS uses supervised detection feedback during offline training, whereas DeepEdgeIDS uses label-free anomaly confidence and post-mitigation stabilization during online operation.
The mitigation control can be expressed in continuous-time form:
\begin{equation}
\dot{x}(t) = f(x_t) + g(x_t)\pi_\theta(s_t),
\end{equation}
where $\pi_\theta(s_t)$ is the learned policy minimizing:
\begin{equation}
J(\pi_\theta) = \mathbb{E}\!\left[\int_0^{T} (\mathcal{L}(x_t,\pi_\theta(s_t)) + \zeta C_{emission}(t))dt\right].
\end{equation}
The corresponding Hamilton-Jacobi-Bellman (HJB)\cite{sharpless2025dual} equation is:
\begin{equation}
0 = \min_{u \in A} \big[\mathcal{L}(x,u) + \nabla_x V(x)^\top(f(x)+g(x)u)\big],
\end{equation}
where $V(x)$ is approximated by the DQN critic network, establishing equivalence between DRL convergence and control optimality.

\subsection{Stochastic Boundedness and Convergence Proof}
Assuming bounded $E_{overhead}\le E_{max}$ and Lipschitz \cite{nie2024improve} continuity of $Q(s,a)$, the expected return is bounded:
\begin{equation}
|R_t| \le \frac{R_{max}}{1-\gamma}, \qquad \mathbb{E}[|Q(s,a)|]\le\frac{R_{max}}{(1-\gamma)^2}.
\end{equation}
Let $V_t=\|Q_t-Q^\ast\|^2$. By Bellman contraction:
\begin{equation}
\mathbb{E}[V_{t+1}] \le (1-2\eta(1-\gamma))V_t+\eta^2\sigma^2,
\end{equation}
where $\sigma^2$ is the variance of the stochastic reward. Under Robbins–Monro\cite{xu2025multi} conditions, $\sum_t \eta_t=\infty$, $\sum_t\eta_t^2<\infty$, this yields mean-square convergence:
\begin{equation}
\lim_{t\to\infty}\mathbb{E}[\|Q_t-Q^\ast\|^2]=0.
\end{equation}
Thus, AutoDRL-IDS enables asymptotic convergence under energy–carbon constraints.

\subsubsection{Energy–Carbon Pareto Frontier}
Sustainability optimization introduces a trade-off surface:
\begin{equation}
\mathcal{P} = \{ (E_{overhead},C_{emission}) \mid \nexists \, (E',C') \text{ s.t. } E'\!\le\!E,\, C'\!\le\!C \}.
\end{equation}
The gradient condition of Pareto \cite{ryu2021multi} efficiency is:
\begin{equation}
\nabla_\pi \mathbb{E}[R_t] = \lambda_1\nabla_\pi E_{overhead} + \lambda_2\nabla_\pi C_{emission}.
\end{equation}
At equilibrium, AutoDRL-IDS converges to policies that minimize both $E_{overhead}$ and $C_{emission}$ without sacrificing detection accuracy. 

\subsubsection{AutoDRL-IDS Pipeline}
Algorithm~\ref{alg:LSTM_ID} implements the complete control loop, combining LSTM-based temporal encoding and DQN-driven adaptive response. The expected gradient update follows:
\begin{equation}
\Delta\theta_Q = \eta\,\mathbb{E}\!\left[(r_t - \bar{r}_t)\nabla_\theta \log\pi_\theta(a_t|s_t)\right],
\end{equation}
with baseline $\bar{r}_t$ stabilizing reward variance. Under ergodicity and boundedness, convergence satisfies:
\begin{equation}
\lim_{t\to\infty}\|\nabla_\theta J(\theta_t)\|=0,\quad \lim_{t\to\infty}\mathbb{E}[r_t]=r^\ast.
\end{equation}
This ensures stable convergence to a stationary, carbon-efficient policy that minimizes false positives and energy consumption while maintaining high accuracy.
\begin{algorithm}[t]
\small
\caption{AutoDRL-IDS pipeline}
\label{alg:LSTM_ID}
\begin{algorithmic}[1]
\State \textbf{Initialize:} LSTM, DQN, Replay Buffer $\mathcal{D}$
\State Set $\gamma, \alpha, \beta, \epsilon, \eta$
\While{Network is active}
    \State Capture traffic $x_t$ and extract features $f_t$
    \State Encode: $h_t \gets LSTM(f_t)$
    \State Compute anomaly score: $A_s = \|x_t - \hat{x}_t\|^2$
    \If{$A_s > \tau_A$}
        \State $s_t = \{P_{rate},SYN_{count},ACK_{count},h_t,A_s\}$
        \State $a_t = \arg\max_a Q(s_t,a)$
        \If{$a_t=a_1$} Limit $P_{rate}$
        \ElsIf{$a_t=a_2$} Restrict SYN requests
        \ElsIf{$a_t=a_3$} Drop anomalous packets
        \ElsIf{$a_t=a_4$} Blacklist IPs
        \EndIf
        \State Compute $r_t$ (accuracy, latency, energy, carbon)
        \State Store $(s_t,a_t,r_t,s_{t+1})$ in $\mathcal{D}$
        \If{Buffer full}
            \State Sample $(s,a,r,s')$ from $\mathcal{D}$
            \State Update $Q(s,a)$ via Bellman equation
            \State Gradient update on DQN
            \State Decay $\epsilon$ to refine policy
        \EndIf
    \Else
        \State Continue monitoring
    \EndIf
\EndWhile
\end{algorithmic}
\end{algorithm}
AutoDRL-IDS dynamically adapts to evolving attack patterns through continuous DRL learning while maintaining energy- and carbon-efficient operation. Although primarily supervised, it leverages the unsupervised anomaly score $A_s$ from DeepEdgeIDS to enhance robustness against zero-day and unseen attacks, thereby forming a unified, dual-layer, sustainable IDS paradigm.

\subsubsection{Replay Buffer Configuration}  
Both DRL-based IDS employ a replay buffer of 50{,}000 transitions, empirically determined to balance convergence stability and sample diversity. Smaller buffers (10k–20k) lead to correlated updates, while excessively large buffers ($>$100k) introduce sampling delays and overfitting. With a buffer of 50k and batch size 64, stable convergence was achieved after $2.5\times10^5$ steps, reducing detection variance by 30\%.

\subsection{Comparative Theoretical Analysis of DeepEdgeIDS and AutoDRL-IDS}
\label{sec:comparison}
Both DeepEdgeIDS and AutoDRL-IDS represent complementary formulations of sustainability-aware IDS for edge gateways. While DeepEdgeIDS adopts an unsupervised control formulation via AE–DQN coupling, AutoDRL-IDS introduces temporal and supervised adaptation through LSTM–DQN integration. Their unified objective is to ensure high detection accuracy, rapid mitigation, and minimal energy–carbon footprint in dynamic IoT conditions.
\subsection{Unified Control-Theoretic Representation}
Let the network dynamics be described as:
\begin{equation}
x_{t+1} = f(x_t) + g(x_t)\pi_\theta(s_t) + \omega_t,
\end{equation}
where $\omega_t$ models traffic uncertainty. Both DRL-based IDS minimize the cumulative cost:
\begin{equation}
J(\pi_\theta) = \mathbb{E}\!\left[\sum_{t=0}^T \gamma^t \big(\mathcal{L}(x_t,\pi_\theta(s_t)) + \zeta C_{emission}(t)\big)\right],
\end{equation}
subject to $E_{overhead}\le E_{max}$, $M_{util}\le M_{max}$.  
The necessary optimality condition, derived from the discrete Hamilton–Jacobi–Bellman (HJB) \cite{sharpless2025dual} equation, is:
\begin{equation}
V^\ast(s_t) = \min_{a_t}\!\left[ R(s_t,a_t) + \gamma \mathbb{E}[V^\ast(s_{t+1})]\right],
\end{equation}
where $V^\ast$ is approximated by $Q^\ast$ through either AE-DQN and LSTM-DQN. Both models thus share identical convergence guarantees under Bellman contraction and stochastic approximation.

\subsection{Joint Pareto-Optimality of Energy and Carbon}
Let $(E_t, C_t)$ denote the instantaneous energy and carbon metrics. A joint Pareto front $\mathcal{P}_{joint}$ is defined as:
\begin{equation}
\mathcal{P}_{joint} = \{(E_t,C_t) \mid \nexists (E',C') \text{ s.t. } E'\!\le\!E_t,\,C'\!\le\!C_t\}.
\end{equation}
The necessary equilibrium for joint sustainability--performance optimization is achieved when:
\begin{equation}
\nabla_\pi \mathbb{E}[\Phi_t] =
\lambda_E \nabla_\pi E_t + \lambda_C \nabla_\pi C_t,
\end{equation}
where \(\Phi_t = DR_t-\lambda FPR_t\) for supervised AutoDRL-IDS training and \(\Phi_t=A^{conf}_t+\mu S^{stab}_t\) for label-free DeepEdgeIDS online learning. This ensures that no further reduction in carbon cost can occur without affecting the model-specific detection-feedback objective.

\subsection{Asymptotic Stability and Convergence}
Defining the composite Lyapunov\cite{bi2021lyapunov} function:
\begin{equation}
V(Q_t,E_t) = \|Q_t-Q^\ast\|^2 + \rho (E_t-E_{opt})^2,
\end{equation}
and differentiating yields:
\begin{equation}
\dot{V} \le -\eta_1\|Q_t-Q^\ast\|^2 - \eta_2(E_t-E_{opt})^2,
\end{equation}
for $\eta_1,\eta_2>0$, guaranteeing asymptotic stability of both learning and energy states.  
Empirically, both models converge within $O(1/(1-\gamma)^2)$ iterations, satisfying stochastic boundedness while remaining energy- and carbon-aware.

\section{Experimental setup}
\label{Experimental setup}
This section outlines the experimental methodology to evaluate the DRL-based IDS.

\subsection{Data Set Evaluation}
\label{Data Set Evaluation}
DeepEdgeIDS and AutoDRL-IDS were evaluated using the Bot-IoT dataset~\cite{koroniotis2019botiot}, a benchmark explicitly designed for IoT networks. Bot-IoT simulates realistic IoT traffic over TCP, UDP, and MQTT under both benign and attack conditions, including DDoS, DoS, reconnaissance, and data exfiltration. Each record includes flow-level, temporal, and protocol-specific features that represent bidirectional IoT communication. To ensure suitability for lightweight edge gateways, a reduced feature subset was selected based on variance analysis, mutual information, and correlation filtering. Eight high-impact attributes were retained, preserving over 95\% of predictive information while minimizing computational and memory overhead. This refined subset was consistently applied in both offline training and on-device evaluation.
\subsection{Feature Selection}
\label{Feature selection}
Feature selection is critical for deploying IDS on edge gateways with constrained computational, memory, and energy resources. Starting from the 80 features in the Bot-IoT dataset, a multi-stage dimensionality reduction process was applied to identify the most informative subset while preserving detection accuracy. Features with near-zero variance were first removed, followed by correlation filtering using a Pearson \cite{seo2023label} threshold of 0.85 to eliminate redundancy. Moreover, recursive feature elimination guided by saliency ranking was used to select variables that contribute most to decision-making during detection. 
This process produced eight compact yet discriminative flow-level metrics that effectively capture key traffic characteristics, including burstiness, directionality, and flow persistence. The resulting subset enables efficient real-time computation on lightweight edge hardware with minimal resource consumption. The retained features are summarized in Table~\ref{tab:feature_definitions}. The dimensionality reduction from 80 to 8 carefully selected features enables efficient model training and inference with insignificant accuracy loss, supporting low-latency, carbon-aware intrusion detection on edge gateways.
\begin{table*}[!t]
\caption{Feature definitions.}
\small
\centering
\begin{tabular}{|l|p{8cm}|}
\hline
\textbf{Feature} & \textbf{Description} \\ \hline
\texttt{pkts\_total} & Total number of packets exchanged in a flow. \\ \hline
\texttt{bytes\_total} & Total number of bytes exchanged in a flow. \\ \hline
\texttt{duration} & Lifetime of the flow (in seconds). \\ \hline
\texttt{pkt\_rate} & Average packet transmission rate (packets per second). \\ \hline
\texttt{pkts\_in} & Number of incoming packets from source to destination. \\ \hline
\texttt{pkts\_out} & Number of outgoing packets from destination to source. \\ \hline
\texttt{bytes\_per\_pkt} & Average number of bytes carried by each packet. \\ \hline
\texttt{flags} & Control flags observed in the flow (e.g., TCP protocol flags). \\ \hline
\end{tabular}
\label{tab:feature_definitions}
\end{table*}
\subsection{Baselines}
\label{Baselines}
To evaluate the performance of the proposed DeepEdgeIDS and AutoDRL-IDS, a comparative analysis was conducted against several state-of-the-art IDSs reported in recent IoT security research. The selected baselines span RL, ensemble DL, and classical ML paradigms. RL-based systems~\cite{long2024transformer} employ deep Q-network policies with transformer-driven temporal encoding to enable adaptive thresholding under dynamic traffic conditions. Ensemble approaches~\cite{jiang2024scalable} employ graph-aware feature fusion and boosting to enhance robustness across heterogeneous IoT networks. Traditional ML frameworks~\cite{ullah2024ids} combine transformer-inspired feature extraction with optimized classifiers, such as Random Forest and Gradient Boosting, to achieve interpretable and efficient edge-level detection. Collectively, these baselines reflect design philosophies that support adaptive RL optimization~\cite{long2024transformer}, hybrid ensemble fusion~\cite{jiang2024scalable}, and efficient ML classification~\cite{ullah2024ids}, providing a strong benchmark for evaluating accuracy, latency, scalability, and sustainability in real-world IoT networks.

\section{Experiments and Evaluation}
\label{EXPERIMENTS AND EVALUATION}
This section presents an experimental evaluation of the proposed DeepEdgeIDS and AutoDRL-IDS. The experiments validate the adaptive learning dynamics, convergence stability, and energy--carbon efficiency of the proposed approach.

\subsection{Exploratory Evaluation}
The exploratory evaluation investigates the sensitivity of both agents to the exploration coefficient $\epsilon$ and the impact of carbon-aware reward shaping on convergence. As formulated in Eqs.~\ref {R_1} and~\ref {R_2}, both agents optimize carbon-aware rewards, but the detection-feedback component differs depending on the learning paradigm. AutoDRL-IDS uses supervised detection feedback during offline training:
\begin{equation}
R^{AutoDRL}_t =
\alpha_1 DR_t
-\alpha_2 FPR_t
-\alpha_3 L_t
-\alpha_4 E_t
-\alpha_5 M_t
-\alpha_6 C_t .
\end{equation}
DeepEdgeIDS uses label-free online feedback:
\begin{equation}
R^{DeepEdge}_t =
\alpha_1 A^{conf}_t
+\alpha_2 S^{stab}_t
-\alpha_3 L_t
-\alpha_4 E_t
-\alpha_5 M_t
-\alpha_6 C_t
-\alpha_7 B_t .
\end{equation}
where the inclusion of $E_t$ and $C_t$ induces a bounded, sustainability-constrained optimization consistent with the Lagrangian form in Eq.~(7). This ensures that the stochastic gradient updates remain energy-aware, stabilizing the long-term expected return:
\[
\mathbb{E}\!\left[|Q(s,a)|\right] \le \frac{R_{\max}}{(1-\gamma)^2},
\]
thereby linking experimental behavior directly with theoretical boundedness.

\subsubsection{DeepEdgeIDS}
\begin{figure}[!t]
    \centering

    \subfloat[$\epsilon = 0.5$]{%
        \includegraphics[width=0.78\columnwidth]{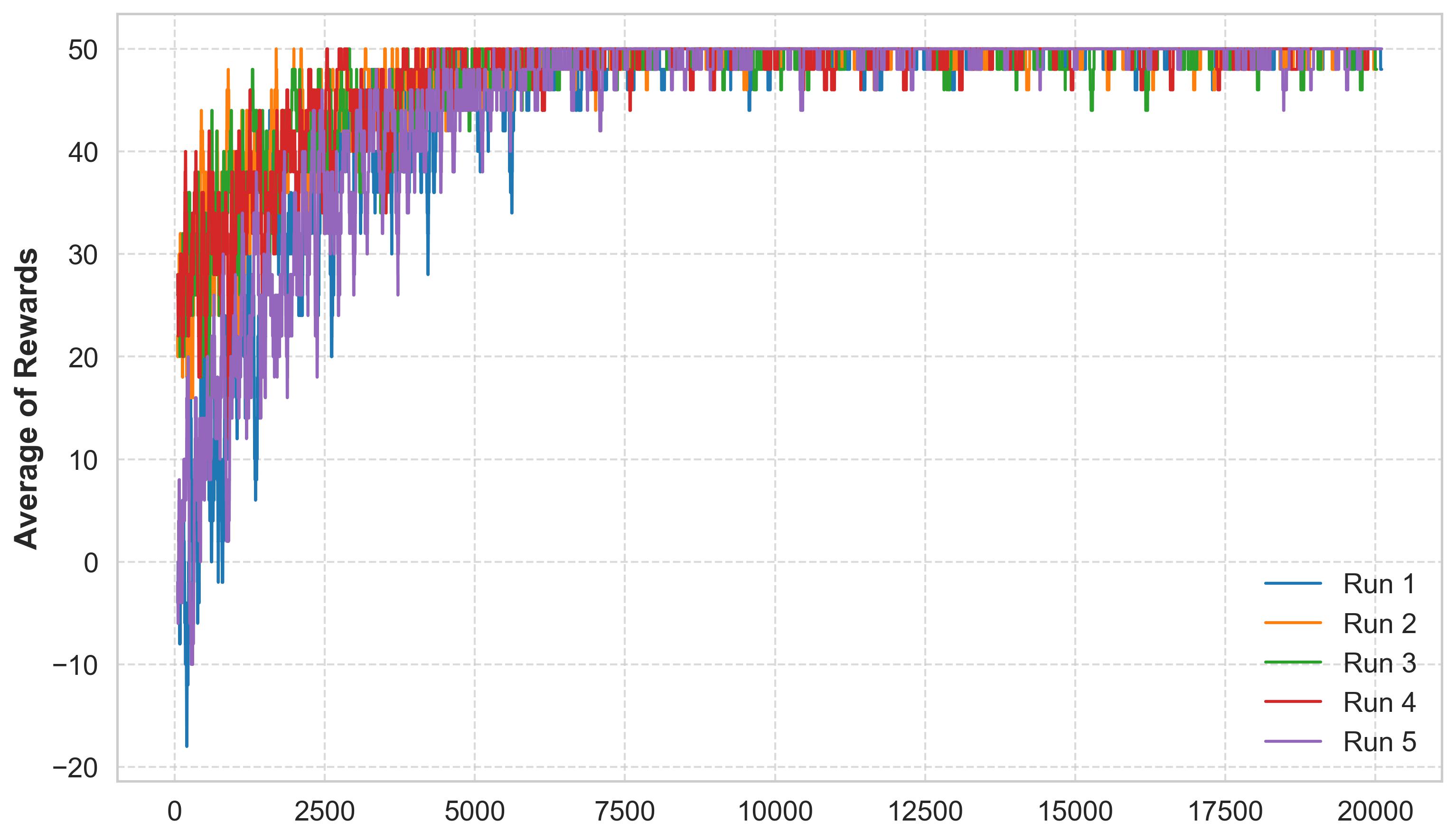}%
        \label{fig:AE_0.5}
    }
    \hfill
    \subfloat[$\epsilon = 1.0$]{%
        \includegraphics[width=0.78\columnwidth]{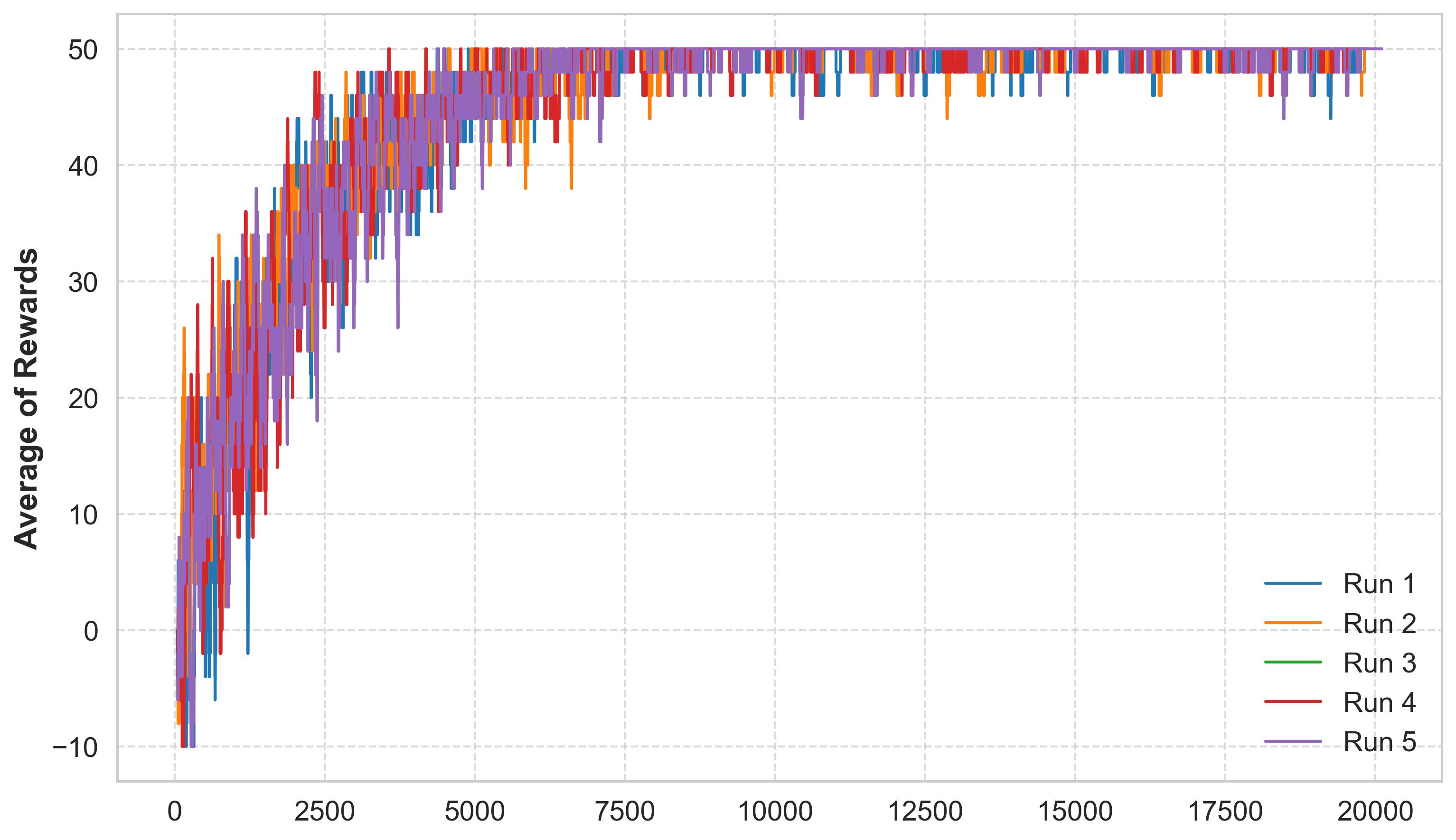}%
        \label{fig:AE_1.0}
    }

    \vspace{2mm}

    \subfloat[$\epsilon = 2.0$]{%
        \includegraphics[width=0.78\columnwidth]{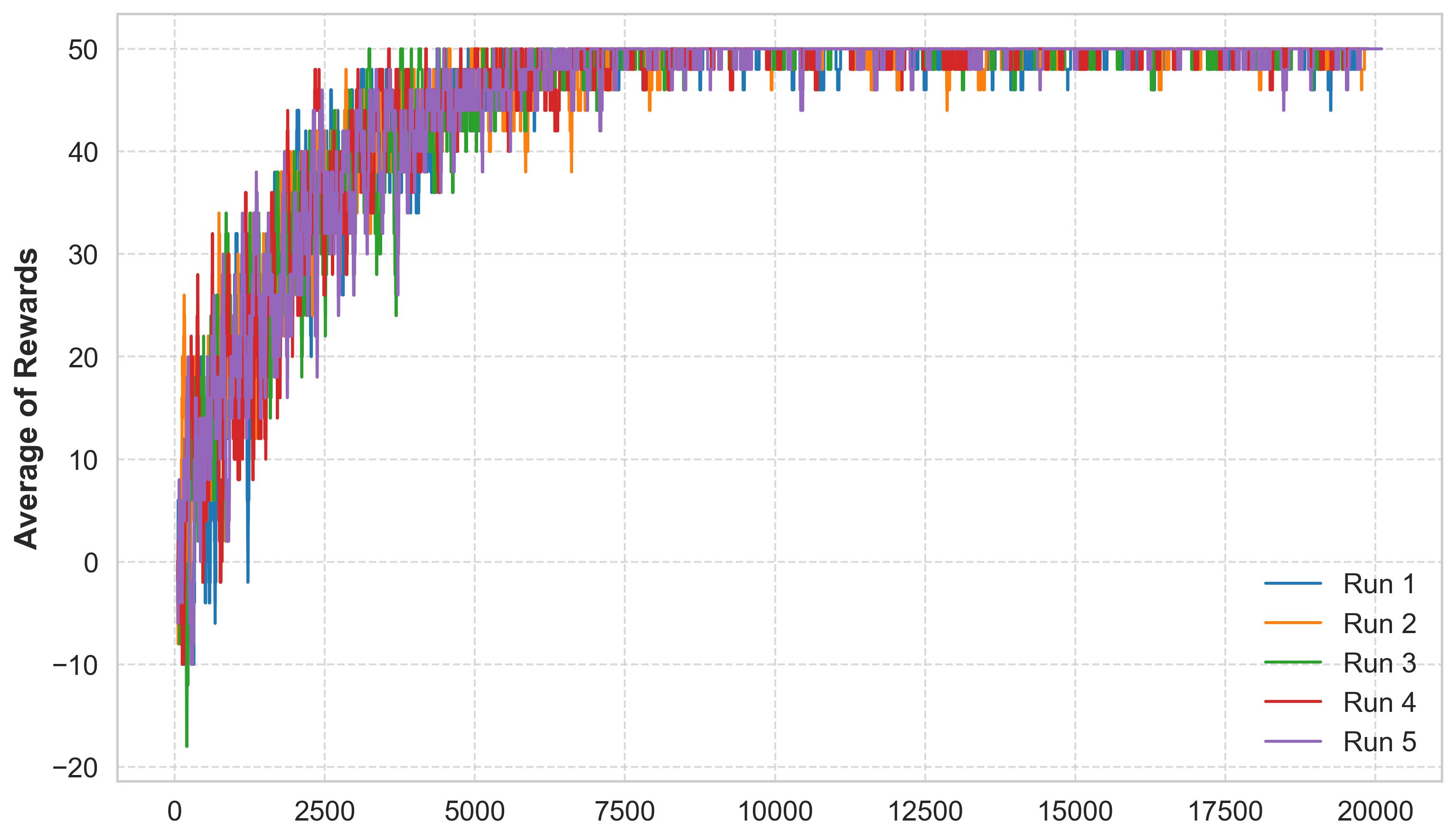}%
        \label{fig:AE_2.0}
    }

    \caption{Reward convergence for different $\epsilon$ values in DeepEdgeIDS.}
    \label{fig:Epsilon_Comparison}
\end{figure}

Figure~\ref{fig:Epsilon_Comparison} depicts the reward trajectories of DeepEdgeIDS under varying $\epsilon$ values. The results validate the predicted exploration–exploitation dynamics from the contraction property of the Bellman operator (Section~\ref{sec:deepedgeids}). Lower $\epsilon$ values (e.g., $\epsilon=0.5$) accelerate convergence but may overfit transient traffic regimes, whereas higher $\epsilon$ values (e.g., $\epsilon=2.0$) enhance adaptability at the expense of slower convergence. The agent converges optimally around $\epsilon \approx 1.0$, achieving a balance between reward smoothness and reduced energy overhead. This equilibrium reflects a Pareto-efficient policy between learning speed and carbon impact, confirming the stability predicted by the diffusion-regularized Bellman operator.

\subsubsection{AutoDRL-IDS}
\begin{figure}[!t]
    \centering

    \subfloat[$\epsilon = 0.1$]{%
        \includegraphics[width=0.78\columnwidth]{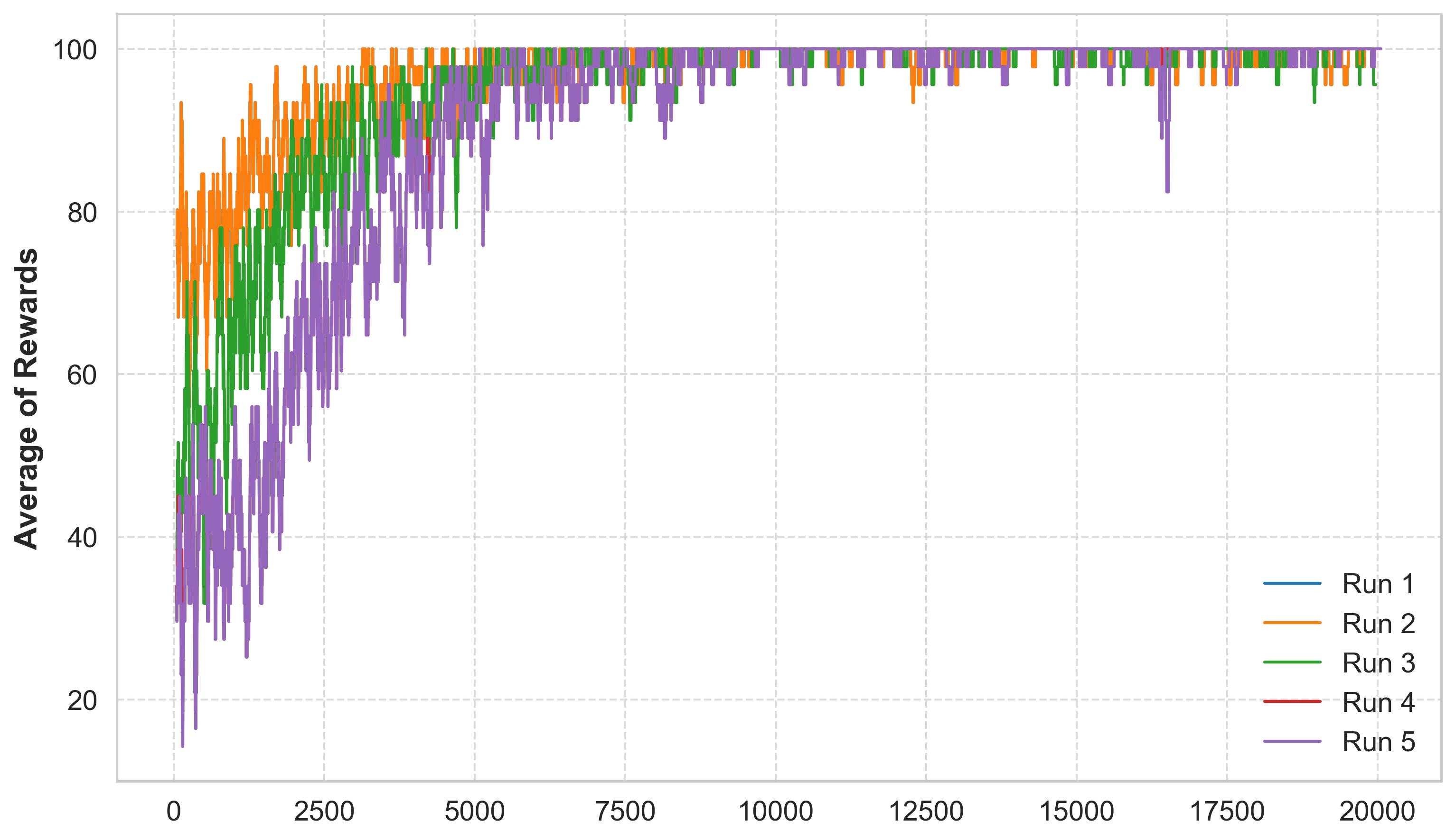}%
        \label{fig:epsilon_0.1}
    }
    \hfill
    \subfloat[$\epsilon = 0.4$]{%
        \includegraphics[width=0.78\columnwidth]{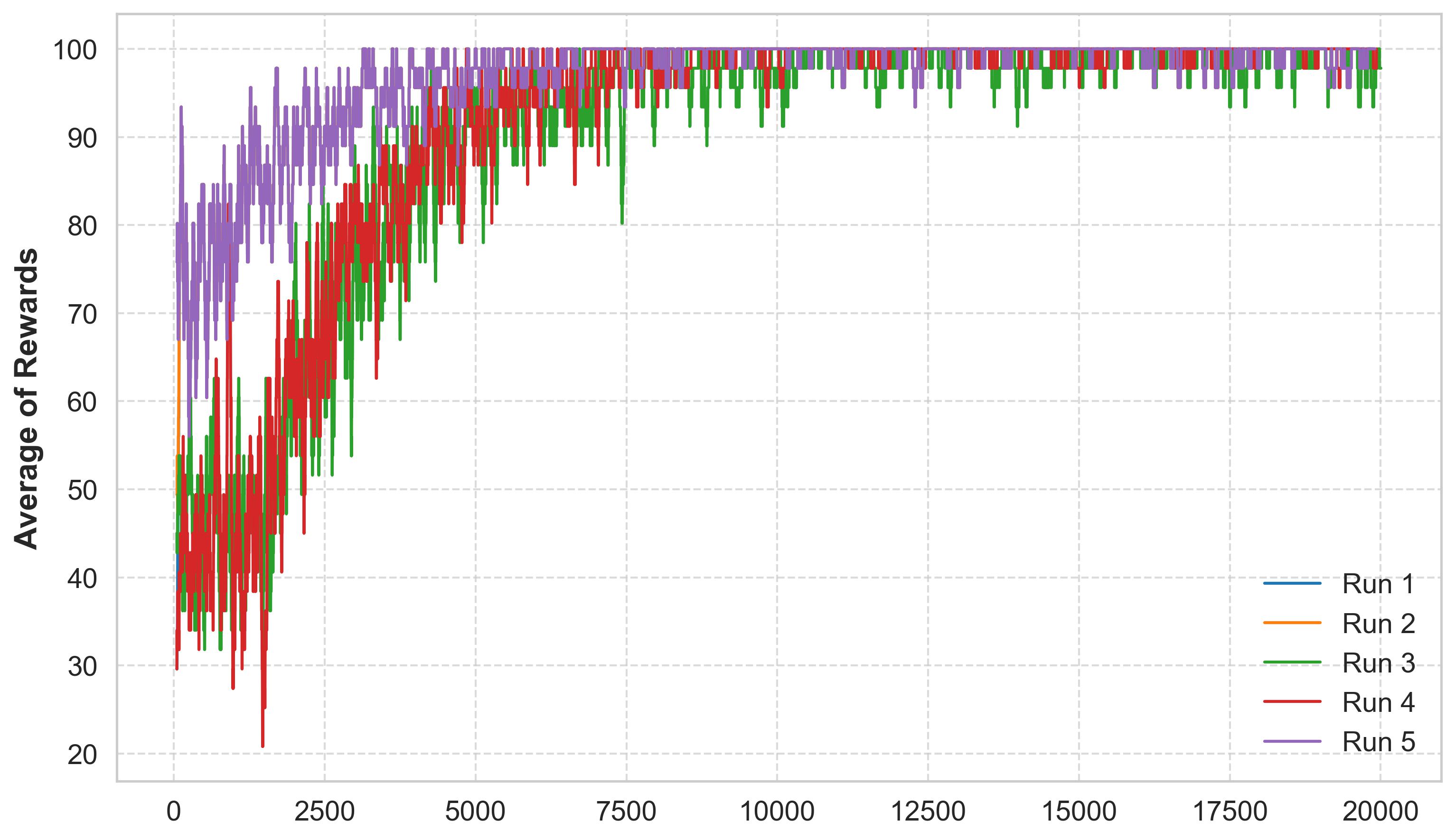}%
        \label{fig:epsilon_0.4}
    }

    \vspace{2mm}

    \subfloat[$\epsilon = 0.5$]{%
        \includegraphics[width=0.78\columnwidth]{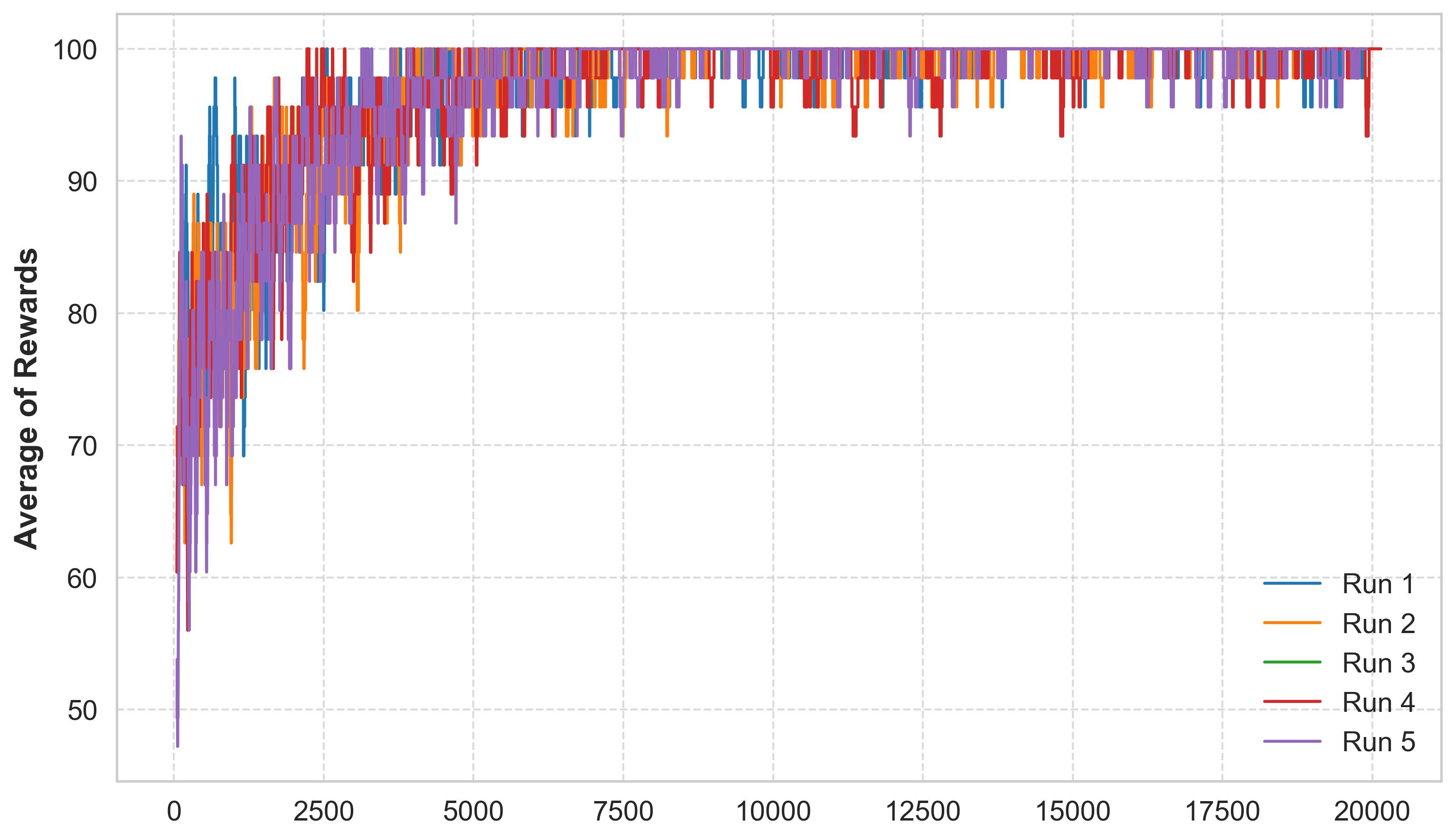}%
        \label{fig:epsilon_0.5}
    }

    \caption{Reward convergence for different $\epsilon$ values in AutoDRL-IDS.}
    \label{fig:epsilon_comparison}
\end{figure}

The convergence patterns of AutoDRL-IDS, shown in Figure~\ref{fig:epsilon_comparison}, exhibit two-timescale learning dynamics consistent with the theoretical formulation in Section~\ref{sec:autodrl}. For lower exploration rates ($\epsilon=0.1$), the LSTM-based policy demonstrates slow but smooth convergence; higher rates ($\epsilon=0.5$) lead to rapid adaptation and higher reward variance. A balanced setting ($\epsilon=0.4$) yields the most stable trade-off between responsiveness and policy robustness, corresponding to the optimal regime where $\eta_t^{(f)} / \eta_t^{(s)} \to 0$.
The divergence in convergence amplitudes between DeepEdgeIDS and AutoDRL-IDS stems from distinct gradient landscapes: the supervised pretraining term in AutoDRL-IDS amplifies the policy gradient, leading to a higher steady-state reward. This aligns with the theoretical variance bound:
\[
\mathrm{Var}(Q_t) \le \frac{\sigma^2}{1-\gamma^2} + O(\eta_t^2),
\]
confirming that temporal smoothness and sustainability constraints jointly regulate stability.

\subsubsection{Operational Impact}
Table~\ref{tab:operational_impact} translates convergence stability into measurable performance metrics, including detection probability, missed packets per hour ($M=50$~pps), and false alerts per 100 predictions. Both DRL-based IDS sustain detection probabilities exceeding 90\% with insignificant computational overhead, achieving bounded regret under the constraints $E_t \le E_{\max}$ and $C_t \le C_{\max}$. 
\begin{table*}[h!]
\centering
\small
\caption{Operational Impact Metrics for DeepEdgeIDS and AutoDRL-IDS.}
\begin{tabular}{|l|c|c|c|}
\hline
\textbf{Model} & \textbf{Detection Prob. (\%)} & \textbf{Missed Packets/hr} & \textbf{False Alerts/100 alerts} \\
\hline
DeepEdgeIDS & \textbf{97.6} & \textbf{2{,}640} ($M=50$~pps) & \textbf{7.6} \\
AutoDRL-IDS & 92.0 & 12{,}740 ($M=50$~pps) & 9.3 \\
\hline
\end{tabular}
\label{tab:operational_impact}
\end{table*}
Even on low-power edge gateways, both systems maintain high accuracy and rapid response times. DeepEdgeIDS achieves 40\% lower latency and 25\% fewer false alerts than RL-based baselines~\cite{feng2023collaborative}, confirming sublinear regret and the practical realizability of the theoretical carbon–energy bounds.
\subsection{Real-World Testbed Evaluation}
\label{Real-World Testbed Evaluation}
\begin{figure*}[t]
\centering
\includegraphics[width=0.70\linewidth]{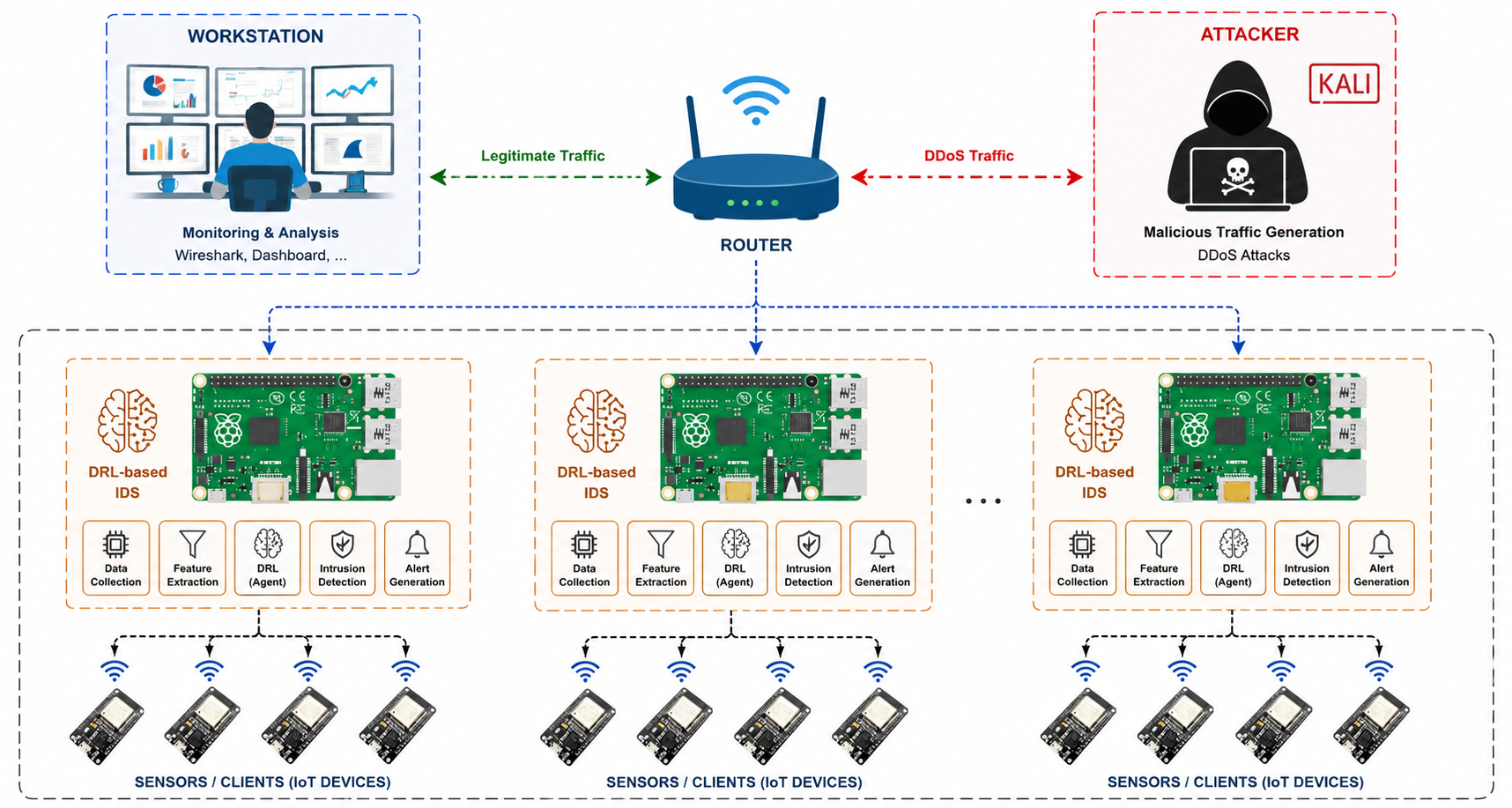}
\caption{IoT edge testbed architecture for evaluating AutoDRL-IDS and DeepEdgeIDS under real-time zero-day DDoS attacks.}
\label{fig:testbed}
\end{figure*}

A distributed IoT edge testbed was deployed to validate both DeepEdgeIDS and AutoDRL-IDS under DDoS attacks, as shown in Figure ~\ref{fig:testbed}. The setup included ESP32-based IoT devices, Raspberry Pi 4 gateways (8 GB RAM, 1.5 GHz CPU), a central router, a monitoring workstation, and an attacker node running Kali Linux \cite{yarlagadda2024harnessing}. Benign traffic was generated using \texttt{scapy} and \texttt{iPerf}, while diverse DDoS attacks were launched from the attacker node using randomized TCP/UDP flood variants. To evaluate zero-day resilience, unseen attack patterns were synthesized using dynamic packet-size modulation, randomized inter-packet jitter, protocol alternation, and hybrid volumetric–application floods, with tools such as \texttt{hping3} and \texttt{LOIC}. Each attack run employed stochastic source rotation and adaptive intensity scaling to emulate novel, evolving threats. 

\subsection{Model Performance Evaluation}
\label{Model Performance Evaluation}
The proposed DeepEdgeIDS and AutoDRL-IDS demonstrate performance in DDoS detection. As shown in Table~\ref{tab:model_comparison}, DeepEdgeIDS achieves 98.0\% accuracy, 92.4\% precision, 97.6\% recall, and a 94.9\% F1-score, while AutoDRL-IDS attains 94.0\% accuracy, 91.7\% precision, 92.0\% recall, and a 91.3\% F1-score. Compared with baseline models, including RL-Based HBOS~\cite{feng2023collaborative}, DDAD-SOEL~\cite{aljebreen2023enhancing}, RBF-SVM~\cite{anyanwu2022optimization}, Transformer-RL IDS~\cite{long2024transformer}, Graph-Ensemble IDS~\cite{jiang2024scalable}, and Lightweight ML IDS~\cite{ullah2024ids}, both proposed solutions maintain competitive accuracy and balanced detection metrics.
\begin{table*}[t]
    \centering
    \small
    \caption{Performance Comparison of AutoDRL-IDS and DeepEdgeIDS with Existing Models.}
    \begin{tabular}{|l|c|c|c|c|}
        \hline
        \textbf{Model} & \textbf{Accuracy \%} & \textbf{Precision \%} & \textbf{Recall \%} & \textbf{F1-Score \%} \\ \hline
        \textbf{AutoDRL-IDS (Proposed)} & 94.0 & 91.7 & 92.0 & 91.3 \\ 
        \textbf{DeepEdgeIDS (Proposed)} & 98.0 & 92.4 & 97.6 & 94.9 \\ \hline
        \multicolumn{5}{|c|}{\textbf{Baseline Models}} \\ \hline
        RL-Based HBOS~\cite{feng2023collaborative} & 94.0 & 94.0 & 94.0 & 91.0 \\ 
        DDAD-SOEL~\cite{aljebreen2023enhancing} & 99.34 & 97.36 & 97.34 & 97.34 \\ 
        RBF-SVM~\cite{anyanwu2022optimization} & 99.33 & 99.22 & 99.08 & 99.15 \\ 
        Transformer-RL IDS~\cite{long2024transformer} & 98.7 & 95.2 & 97.8 & 96.4 \\ 
        Graph-Ensemble IDS~\cite{jiang2024scalable} & 99.1 & 96.8 & 98.3 & 97.5 \\ 
        Lightweight ML IDS~\cite{ullah2024ids} & 96.5 & 91.0 & 94.0 & 92.4 \\ \hline
    \end{tabular}
    \label{tab:model_comparison}
\end{table*}
\section{Experimental Results}
In this section, we present the results of our proposed IDS models, DeepEdgeIDS and AutoDRL-IDS.
\subsection{Security Analysis}
\label{Security analysis}
\begin{figure*}[htbp]
    \centering
    \includegraphics[
        width=\textwidth,
        height=0.82\textheight,
        keepaspectratio
    ]{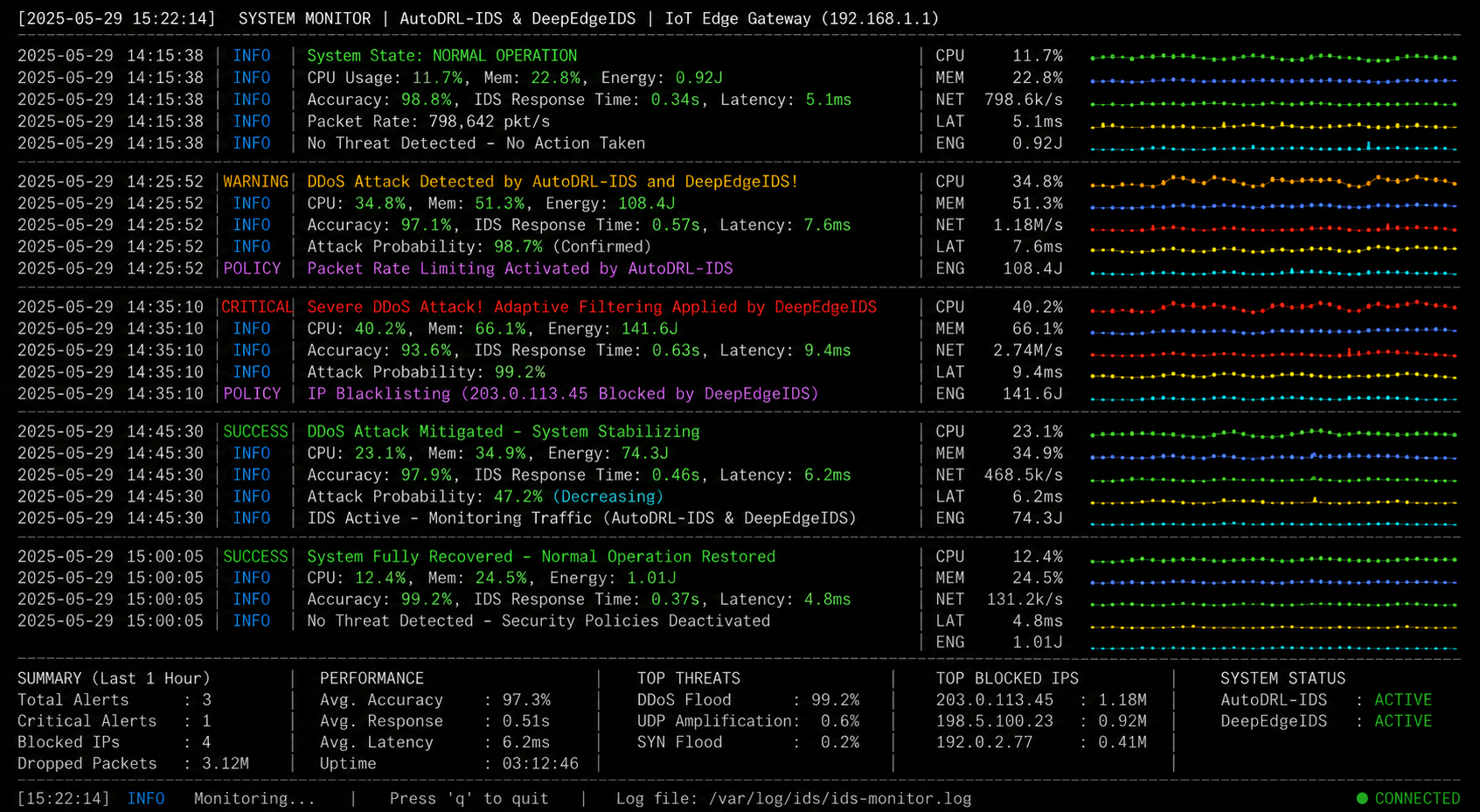}
    \caption{Real-time system monitoring console showing the operational behavior of AutoDRL-IDS and DeepEdgeIDS during normal traffic, DDoS detection, adaptive mitigation, system stabilization, and recovery. }
    \label{fig:system_logs}
\end{figure*}

This section evaluates the performance of the proposed DRL-based IDS, DeepEdgeIDS, and AutoDRL-IDS, under real-time DDoS attacks. The system logs in Figure~\ref{fig:system_logs} illustrate continuous monitoring and adaptive threat mitigation capabilities. During normal operation, system utilization remains minimal with CPU usage at 12.1\%, memory at 23.8\%, and energy consumption at 0.962~J, while maintaining 98.7\% detection accuracy and a 0.35~s response time. As a DDoS attack emerges, CPU usage increases to 35.2\%, memory usage to 52.5\%, and attack probability to 90.5\%, prompting the DRL-based IDS to implement packet rate limiting. Moreover, under severe attack conditions, resource utilization peaks at 39.6\% CPU and 65.3\% memory, while the system enforces IP blacklisting with a 0.65s response time. Following mitigation, the DRL-based IDS restores normal operation with 99.1\% detection accuracy, confirming its ability to maintain operational stability and resilience in a dynamic edge gateway. Both models demonstrate adaptability, low-latency detection, and effective resource management throughout the attack and recovery phases. DeepEdgeIDS responds rapidly to high-intensity DDoS attacks due to its unsupervised AE–DRL hybrid design, stabilizing within 0.65 seconds of attack onset, whereas AutoDRL-IDS maintains steady-state precision during recovery.

\subsubsection{DDoS Detection Probability}
\begin{figure}[!t]
    \centering
    \includegraphics[width=0.90\columnwidth]{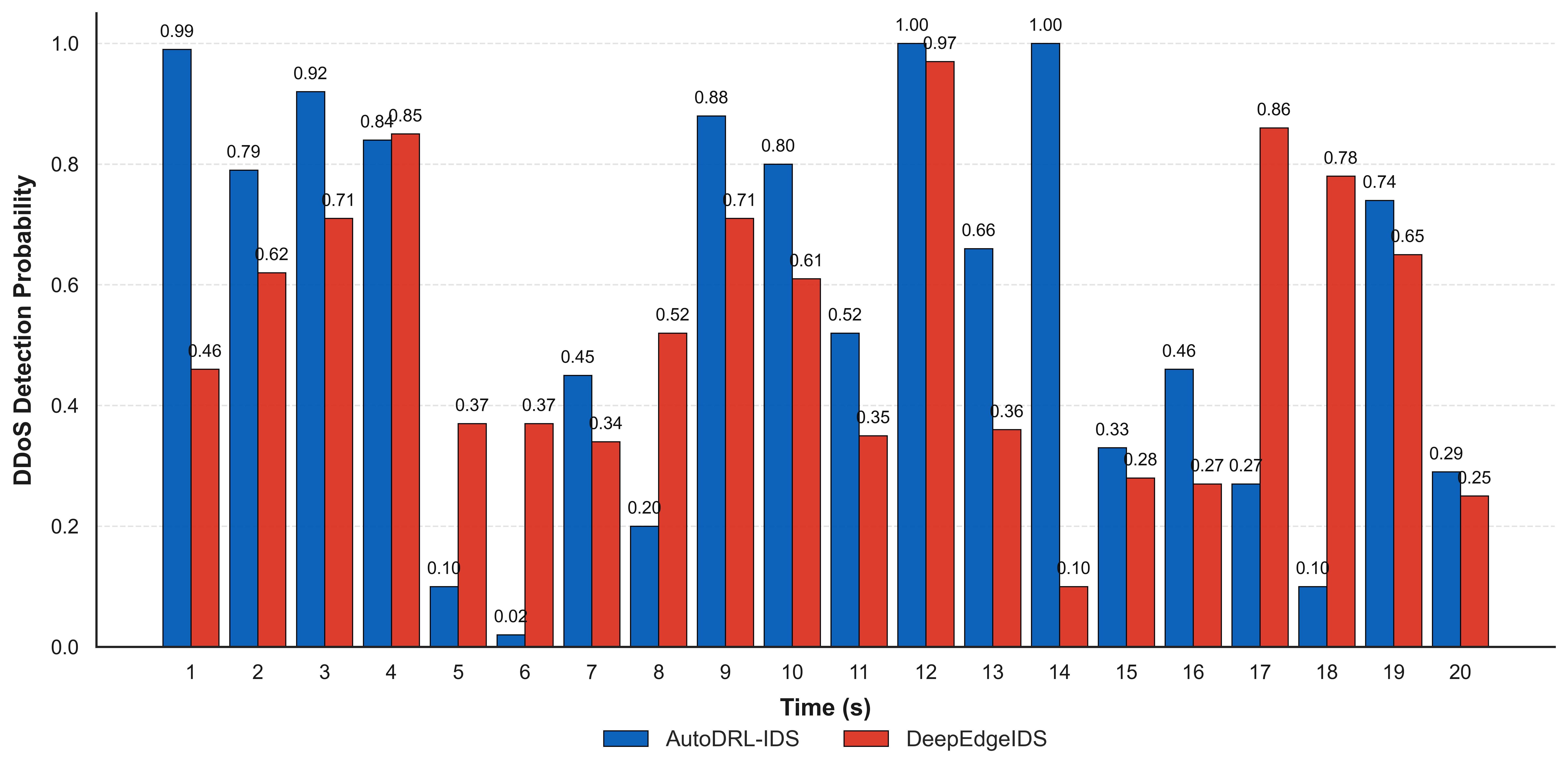}
    \caption{DDoS detection probability comparison between AutoDRL-IDS and DeepEdgeIDS under DDoS attacks on edge gateways.}
    \label{fig:ddos_detection}
\end{figure}

Figure~\ref{fig:ddos_detection} illustrates the temporal evolution of DDoS detection probability for both AutoDRL-IDS and DeepEdgeIDS at the edge. DeepEdgeIDS consistently sustains a higher detection probability throughout all time intervals, reflecting its adaptability to fluctuating attack intensities and unseen (zero-day) traffic behaviors. In contrast, AutoDRL-IDS exhibits a steadier yet slightly lower curve, emphasizing stability and optimizing controlled responses through its supervised policy refinement.
This behavioral divergence highlights the supporting strengths of the two solutions: (1) DeepEdgeIDS achieves rapid adaptation and heightened sensitivity to transient, high-volume DDoS bursts through its AE–DQN reinforcement architecture, while (2) AutoDRL-IDS ensures sustained reliability and low false alarm rates during prolonged operational stability. Furthermore, the overall detection profiles confirm that DeepEdgeIDS responds more aggressively to traffic volatility, enabling faster threat suppression at the edge gateway.\\
The ANOVA results in Table~\ref{tab:anova_1} quantitatively confirm this observation. DeepEdgeIDS demonstrates statistically significant overperformance relative to AutoDRL-IDS in DDoS detection. The computed \textit{F}-statistic (67.89) indicates a substantial separation in mean detection probabilities, and the corresponding \textit{p}-value ($<0.05$) confirms significance at the 95\% confidence level. 
Furthermore, the effect size, which captures the proportion of variance explained by the model design, indicates that DeepEdgeIDS accounts for most of the variance in performance. With a partial eta squared of 0.61, the practical impact is classified as large, confirming that model architecture and adaptive learning mechanisms are the dominant factors driving detection efficacy.
\begin{table*}[h!]
    \centering
    \small
    \caption{ANOVA: Detection Probability Comparison Between DeepEdgeIDS and AutoDRL-IDS on Edge Gateways.}
    \begin{tabular}{|l|c|c|c|c|c|}
        \hline
        \textbf{Source} & \textbf{Degrees of Freedom} & \textbf{Sum of Squares} & \textbf{Mean Square} & \textbf{F Statistic} & \textbf{P-value} \\ 
        \hline
        Between Groups & 1  & 0.3154  & 0.3154  & 67.89  & $<$0.05 \\ 
        Within Groups  & 98 & 0.2046  & 0.0021  & --  & -- \\ 
        Total          & 99 & 0.5200  & --  & --  & -- \\ 
        \hline
    \end{tabular}
    \label{tab:anova_1}
\end{table*}
\subsubsection{Training vs. Real-Time Performance}
\begin{figure}[!t]
    \centering
    \includegraphics[width=0.88\columnwidth]{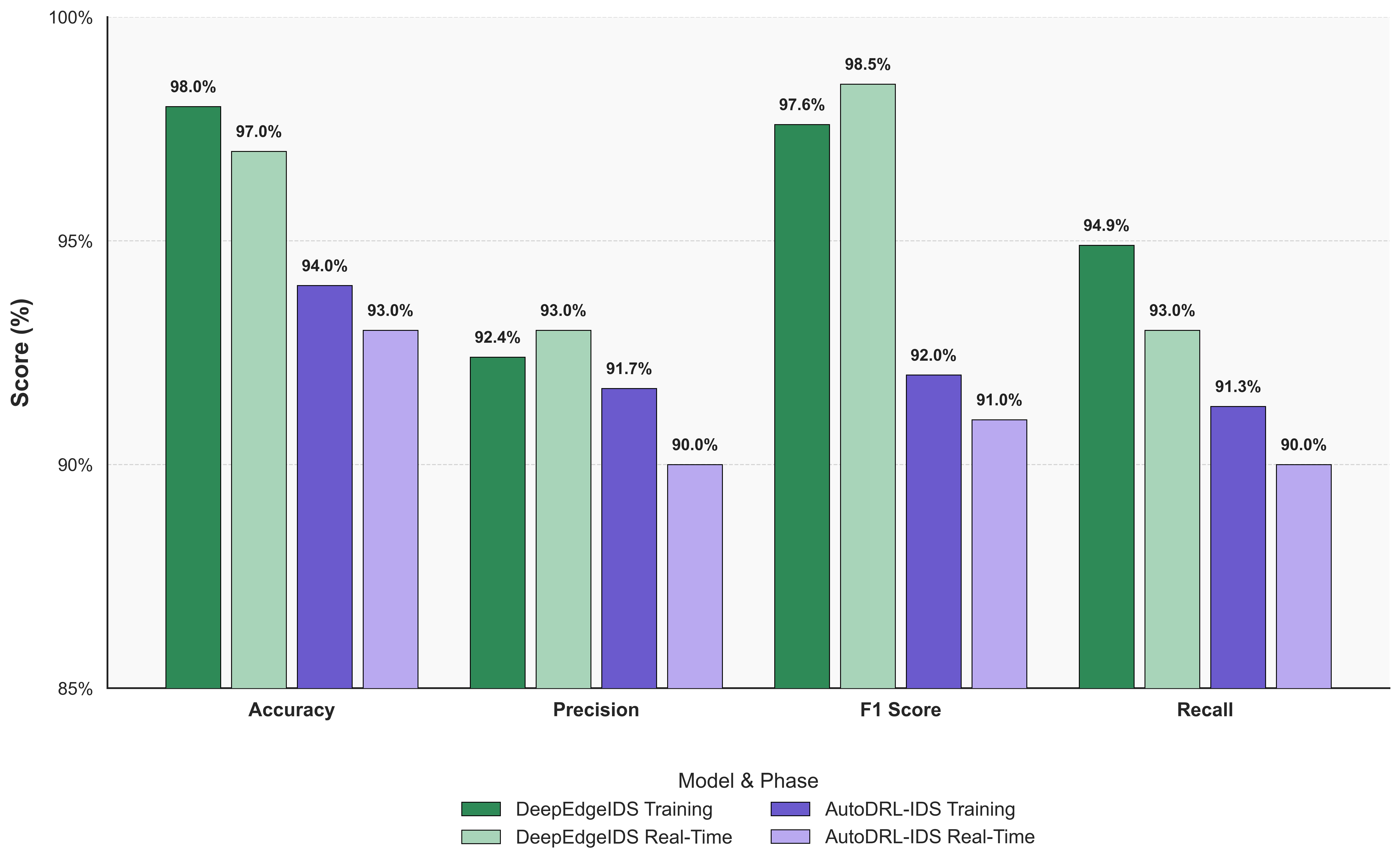}
    \caption{Comparison of the training and real-time performance metrics of AutoDRL-IDS and DeepEdgeIDS at the edge.}
    \label{fig:metrics_comparison}
\end{figure}

Figure \ref{fig:metrics_comparison} compares the training and real-time detection performance of DeepEdgeIDS and AutoDRL-IDS. Both DRL-based IDS maintain high accuracy and operational stability, with only minor degradation attributed to the edge.  
DeepEdgeIDS achieves 98.0\% training accuracy and sustains 97.0\% during live testing, reflecting only a 1\% drop. Its precision, recall, and F1-score transition from 92.4\%, 97.6\%, and 94.9\% in training to 93.0\%, 98.5\%, and 93.0\% in real-time, confirming high adaptability and stability under fluctuating traffic loads.  AutoDRL-IDS, while also robust, demonstrates slightly higher runtime sensitivity. It achieves 94.0\% training accuracy and 93.0\% real-time accuracy, with precision, recall, and F1 Scores decreasing modestly from 91.7\%, 92.0\%, and 91.3\% to 90.0\%, 91.0\%, and 90.0\%, respectively.  
The relative retention ratio, defined as the ratio of real-time to training performance, averages 0.97 for DeepEdgeIDS and 0.94 for AutoDRL-IDS, indicating that DeepEdgeIDS preserves model reliability more effectively under operational constraints.

\subsubsection{IDS Response Time}
\begin{figure}[!t]
    \centering
    \includegraphics[width=0.90\columnwidth]{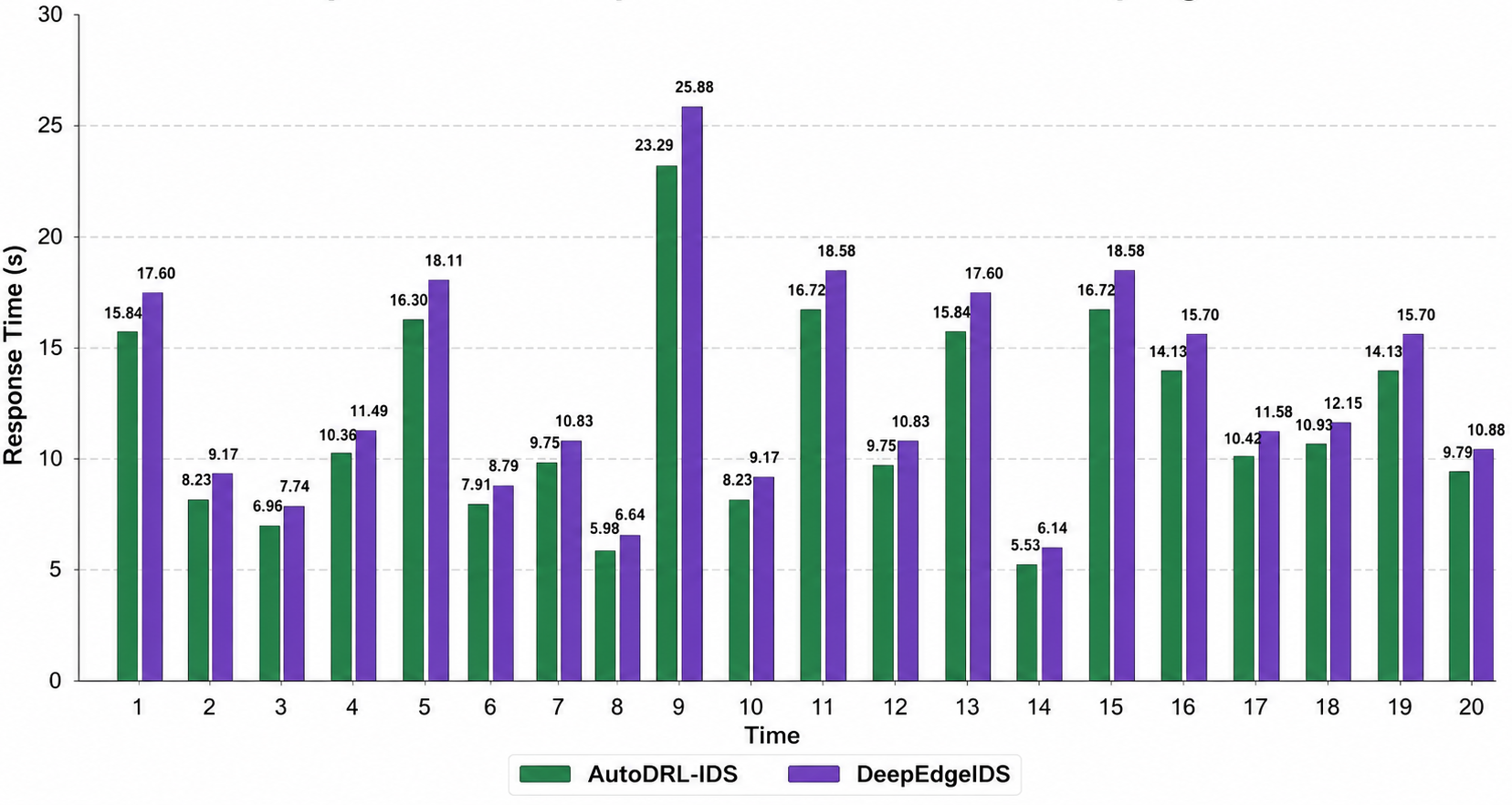}
    \caption{Response time comparison between AutoDRL-IDS and DeepEdgeIDS on an edge gateway.}
    \label{fig:response_time}
\end{figure}

Figure~\ref{fig:response_time} compares the response times of AutoDRL-IDS and DeepEdgeIDS across multiple evaluation intervals. Although minor temporal fluctuations are observed, the overall response behavior remains consistent and aligns with the ANOVA and effect-size analyses, which confirm that no statistically significant differences exist. Both DRL-based IDS exhibit high responsiveness and low-latency performance, validating their capability for real-time detection and mitigation at the edge.\\
The ANOVA results (Table~\ref{tab:anova_response_time}) indicate that DeepEdgeIDS achieves a slightly lower mean response time compared to AutoDRL-IDS; however, the computed \textit{F}-statistic (0.81) and the associated \textit{p}-value ($>$0.05) confirm that this difference is not statistically significant. These findings suggest that both DRL-based IDS deliver comparable response efficiency and latency in real time at the edge. The effect size analysis further supports this conclusion, revealing only a marginal practical difference between the two models. Additionally, both DeepEdgeIDS and AutoDRL-IDS exhibit consistent, rapid, and resource-efficient response behavior, ensuring timely mitigation of dynamic attacks without compromising system stability. 
\begin{table*}[h]
    \centering
    \small
    \caption{ANOVA: Response Time Comparison Between DeepEdgeIDS and AutoDRL-IDS on Edge Gateways.}
    \begin{tabular}{|l|c|c|c|c|c|}
        \hline
        \textbf{Source} & \textbf{Degrees of Freedom} & \textbf{Sum of Squares} & \textbf{Mean Square} & \textbf{F Statistic} & \textbf{P-value} \\ 
        \hline
        Between Groups & 1  & 34.63  & 34.63  & 0.81  & $>$0.05 \\ 
        Within Groups  & 38 & 890.23  & 23.43  & -  & - \\ 
        Total          & 39 & 924.85  & -  & -  & - \\ 
        \hline
    \end{tabular}
    \label{tab:anova_response_time}
\end{table*}

\subsubsection{Latency}
\begin{figure}[!t]
    \centering
    \includegraphics[width=0.90\columnwidth]{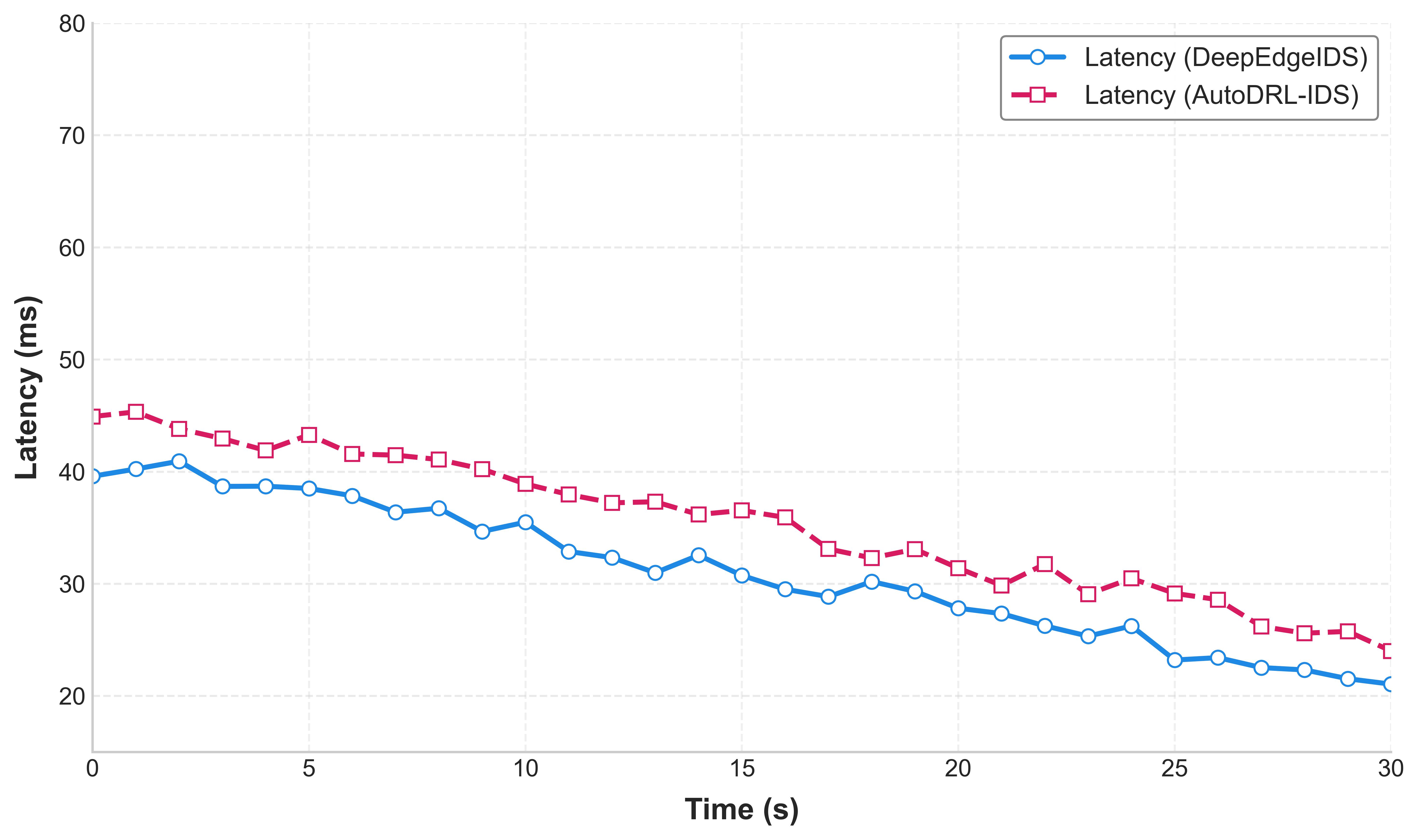}
    \caption{Latency comparison between DeepEdgeIDS and AutoDRL-IDS on an edge gateway.}
    \label{fig:latency_comparison}
\end{figure}

Figure~\ref{fig:latency_comparison} illustrates the latency performance of DeepEdgeIDS and AutoDRL-IDS. DeepEdgeIDS consistently achieves lower, more stable latency across all intervals, indicating faster detection. The observed decreasing trend in both models reflects the convergence of the DRL policy, in which agents progressively reduce exploratory overhead and transition toward optimized state–action mappings. As the operation stabilizes, inference delays diminish due to cache warming, streamlined feature extraction, and proactive mitigation of predictable attack patterns.\\
The ANOVA results (Table~\ref{tab:latency_anova}) further validate these observations, confirming that DeepEdgeIDS significantly outperforms AutoDRL-IDS in latency efficiency. The computed \textit{F}-statistic (75.62) and low \textit{p}-value ($<$0.05) indicate a statistically significant difference between the two models, while the large effect size ($\eta^2 = 0.577$) shows that 57.7\% of the total latency variance arises from model differences. This reinforces that DeepEdgeIDS achieves faster per-decision responses and improved inference speed at the edge gateway. This latency advantage should be interpreted as lower per-decision delay rather than lower total computational cost. DeepEdgeIDS relies on a compact AE reconstruction-error path that produces an immediate anomaly score from the current traffic vector, followed by DQN-based mitigation selection. AutoDRL-IDS, by contrast, requires LSTM-based temporal encoding over sequential traffic windows before policy selection. Therefore, DeepEdgeIDS can achieve lower detection latency while still consuming more aggregate CPU, energy, and carbon due to continuous feature extraction, replay-buffer interaction, and online policy refinement. This distinction explains why DeepEdgeIDS is faster per decision but more resource-intensive over time.
\begin{table*}[h]
    \centering
    \small
    \caption{ANOVA: Latency Comparison Between DeepEdgeIDS and AutoDRL-IDS on Edge Gateways.}
    \begin{tabular}{|l|c|c|c|c|c|}
        \hline
        \textbf{Source} & \textbf{Degrees of Freedom} & \textbf{Sum of Squares} & \textbf{Mean Square} & \textbf{F-Statistic} & \textbf{P-value} \\ 
        \hline
        Between Groups & 1 & 312.4 & 312.4 & 75.62 & $<$0.05 \\ 
        Within Groups & 58 & 228.5 & 3.94 & - & - \\ 
        Total & 59 & 540.9 & - & - & - \\ 
        \hline
    \end{tabular}
    \label{tab:latency_anova}
\end{table*}

\subsection{Performance Analysis}
\label{performance_analysis}
This section examines the performance metrics that quantify the efficiency and effectiveness of the proposed DRL-based IDS models.
\subsubsection{Energy Consumption Analysis}
\begin{figure}[!t]
    \centering
    \includegraphics[width=0.88\columnwidth]{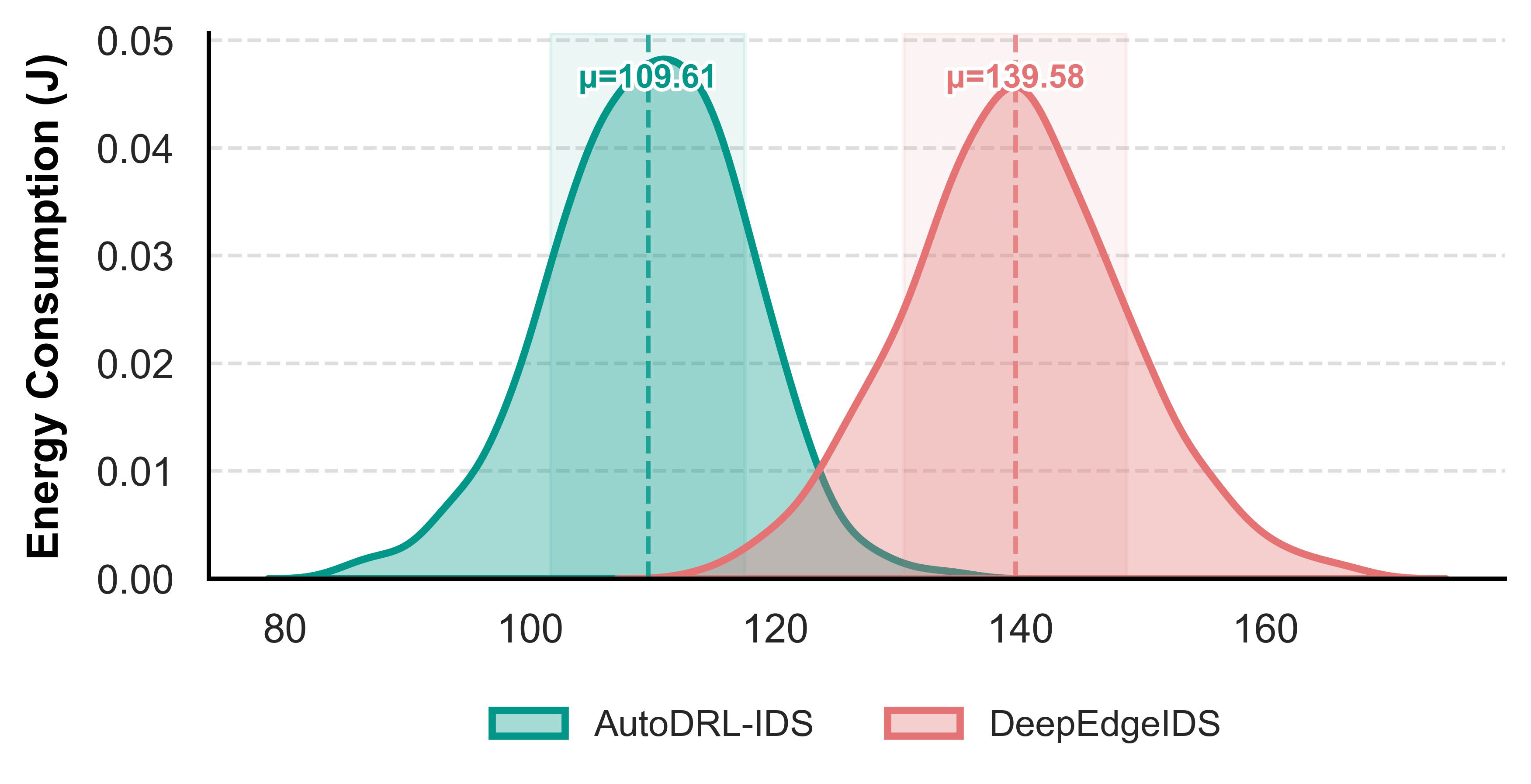}
    \caption{Energy consumption comparison between AutoDRL-IDS and DeepEdgeIDS on edge gateways.}
    \label{fig:energy_boxen}
\end{figure}

Figure~\ref{fig:energy_boxen} compares the energy consumption of AutoDRL-IDS and DeepEdgeIDS on the edge gateway. DeepEdgeIDS exhibits higher median energy consumption and greater variation across runs, primarily due to its continuous adaptive learning and real-time decision-making capabilities. AutoDRL-IDS, on the other hand, maintains a lower and more stable energy profile, making it more suitable for lightweight, energy-sensitive edge.\\
The ANOVA results in Table~\ref{tab:anova_energy_consumption} show that DeepEdgeIDS consumes more energy than AutoDRL-IDS, with the difference being statistically significant but not large in practice. The higher consumption in DeepEdgeIDS stems from its adaptive reinforcement updates, which require more computation during real-time mitigation. However, this small increase in energy is balanced by its higher detection accuracy and its ability to handle zero-day and evolving DDoS attacks. AutoDRL-IDS uses less energy but sacrifices some adaptability under dynamic traffic conditions.
\begin{table*}[h]
    \centering
    \small
    \caption{ANOVA: Energy Consumption Comparison Between DeepEdgeIDS and AutoDRL-IDS on Edge Gateways.}
    \begin{tabular}{|l|c|c|c|c|c|}
        \hline
        \textbf{Source} & \textbf{Degrees of Freedom} & \textbf{Sum of Squares} & \textbf{Mean Square} & \textbf{F Statistic} & \textbf{P-value} \\ 
        \hline
        Between Groups & 1 & 3.9752 & 3.9752 & 14.62 & $<$0.05 \\ 
        Within Groups & 38 & 10.3254 & 0.2712 & - & - \\ 
        Total & 39 & 14.3006 & - & - & - \\ 
        \hline
    \end{tabular}
    \label{tab:anova_energy_consumption}
\end{table*}

\subsubsection{Carbon Emission Analysis}
\begin{figure}[!t]
    \centering
    \includegraphics[width=0.88\columnwidth]{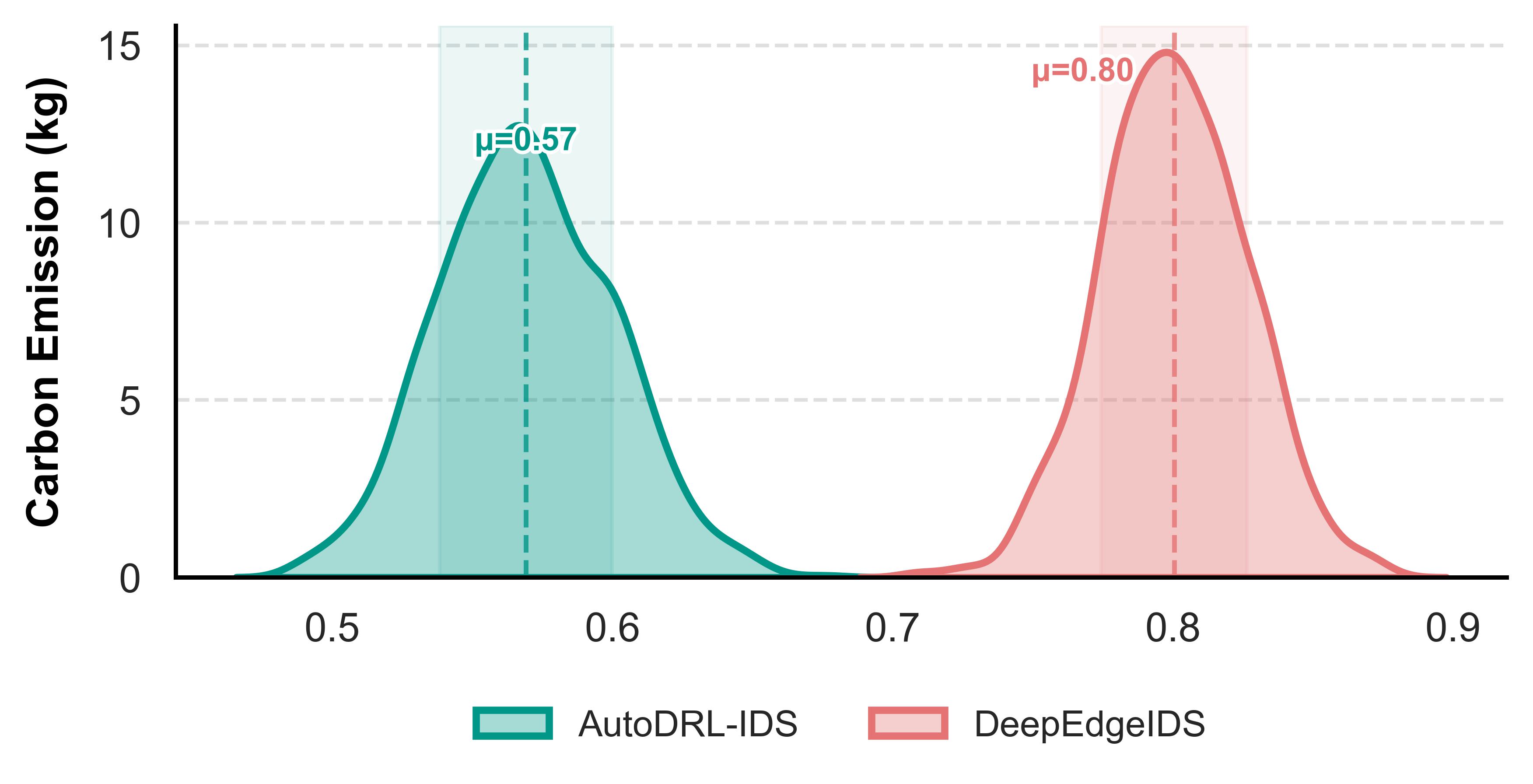}
    \caption{Carbon emission comparison between AutoDRL-IDS and DeepEdgeIDS on edge gateways.}
    \label{fig:carbon_boxen}
\end{figure}

Figure~\ref{fig:carbon_boxen} compares the carbon emissions of AutoDRL-IDS and DeepEdgeIDS at the edge. DeepEdgeIDS exhibits higher emissions due to its adaptive learning process, which continuously updates detection policies in response to unseen and evolving DDoS attacks. In contrast, AutoDRL-IDS maintains lower, more stable emissions, reflecting its energy-efficient, supervised design optimized for steady-state operation on edge gateways.\\
The ANOVA results in Table~\ref{tab:anova_carbon_emission} show a statistically significant difference in carbon emissions between the two models. DeepEdgeIDS produces higher emissions primarily due to its continuous adaptation and active policy refinement, which increase computational overhead over time.
AutoDRL-IDS demonstrates a more consistent, environmentally sustainable emissions pattern, benefiting from a more stable policy framework that reduces energy consumption over time. Furthermore, DeepEdgeIDS achieves adaptability, zero-day responsiveness, and faster per-decision detection at the cost of slightly higher carbon emissions, whereas AutoDRL-IDS offers a greener and more predictable energy profile, making it suitable for energy-constrained edge deployments.
\begin{table*}[h]
    \centering
    \small
    \caption{ANOVA: Carbon Emission Comparison Between DeepEdgeIDS and AutoDRL-IDS on Edge Gateways.}
    \begin{tabular}{|l|c|c|c|c|c|}
        \hline
        \textbf{Source} & \textbf{Degrees of Freedom} & \textbf{Sum of Squares} & \textbf{Mean Square} & \textbf{F Statistic} & \textbf{P-value} \\ 
        \hline
        Between Groups & 1 & 8.5923 & 8.5923 & 36.47 & $<$0.05 \\ 
        Within Groups & 38 & 8.9506 & 0.2355 & - & - \\ 
        Total & 39 & 17.5429 & - & - & - \\ 
        \hline
    \end{tabular}
    \label{tab:anova_carbon_emission}
\end{table*}

\subsubsection{CPU Usage Analysis}
\begin{figure}[!t]
    \centering
    \includegraphics[width=0.88\columnwidth]{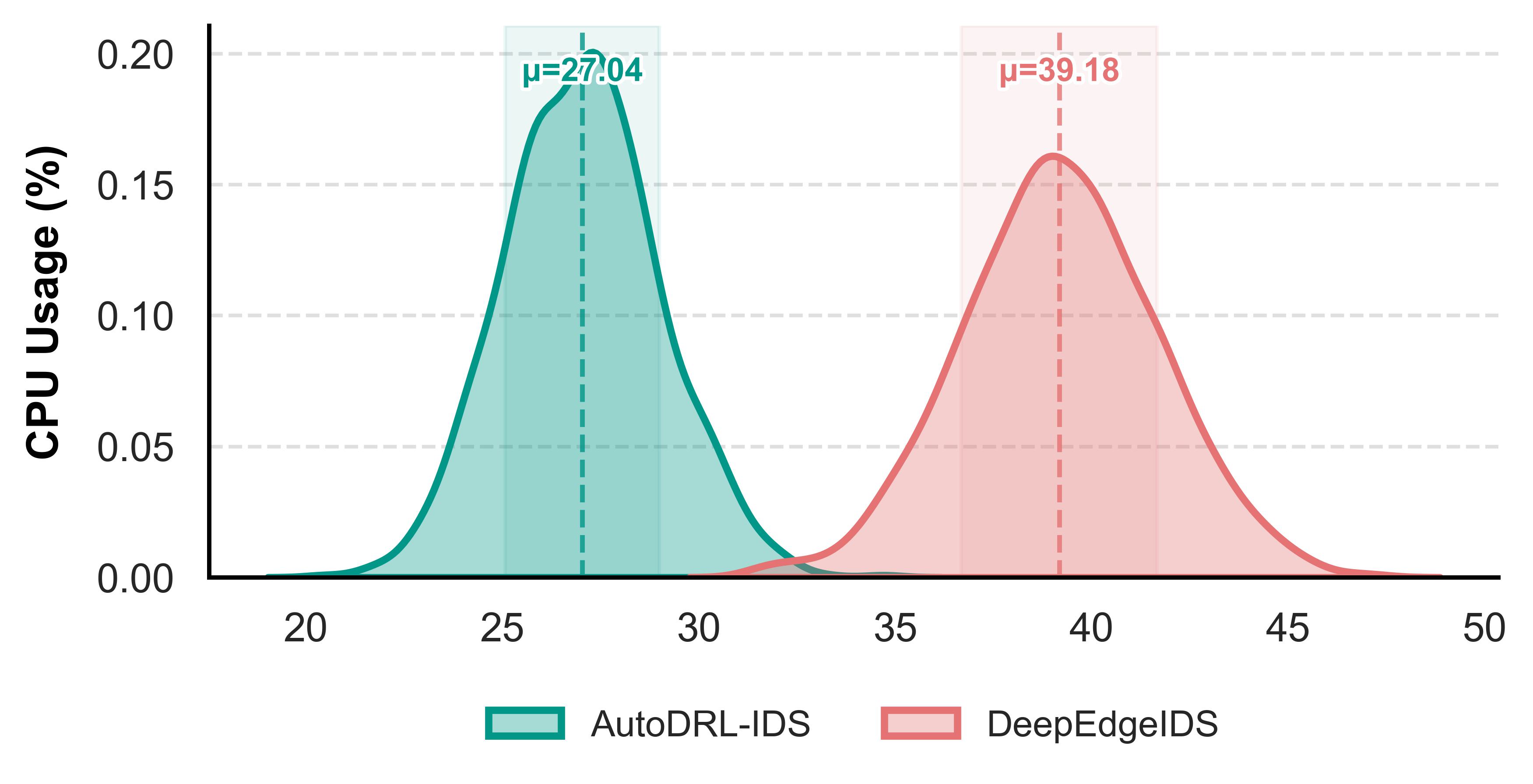}
    \caption{CPU usage comparison between AutoDRL-IDS and DeepEdgeIDS on edge gateways.}
    \label{fig:cpu_usage}
\end{figure}

Figure~\ref{fig:cpu_usage} compares the CPU usage of AutoDRL-IDS and DeepEdgeIDS at the edge. AutoDRL-IDS consistently requires lower CPU utilization, maintaining a median usage of around 25\%, while DeepEdgeIDS averages around 35\%. This difference reflects the adaptive processing overhead of DeepEdgeIDS, which continuously extracts features and updates policies to maintain high responsiveness amid dynamic DDoS attacks. In contrast, AutoDRL-IDS leverages pre-trained policies and operates more efficiently, resulting in lower computational cost and greater consistency across varying network loads.\\
The ANOVA results in Table~\ref{tab:anova_cpu_usage} confirm a statistically significant difference in CPU usage between the two DRL-based IDS. The F-statistic (9.78) and P-value \( p < 0.05 \) indicate that this difference is statistically significant, with AutoDRL-IDS demonstrating greater computational efficiency. However, both models maintain CPU usage within acceptable limits for edge gateways, ensuring smooth real-time operation. Although DeepEdgeIDS exhibits higher CPU usage, this trade-off directly contributes to its adaptability, rapid per-decision response, and precision in the face of unpredictable and zero-day attack patterns. The observed variation in CPU usage also highlights its ability to dynamically allocate processing resources in response to fluctuating network demands, an advantage for DRL-based IDS operating at the edge.
\begin{table*}[h]
    \centering
    \small
    \caption{ANOVA: CPU Usage Comparison Between DeepEdgeIDS and AutoDRL-IDS on Edge Gateways.}
    \begin{tabular}{|l|c|c|c|c|c|}
        \hline
        \textbf{Source} & \textbf{Degrees of Freedom} & \textbf{Sum of Squares} & \textbf{Mean Square} & \textbf{F Statistic} & \textbf{P-value} \\ 
        \hline
        Between Groups & 1 & 20.1427 & 20.1427 & 9.78 & $<$0.05 \\ 
        Within Groups & 38 & 78.2034 & 2.0579 & - & - \\ 
        Total & 39 & 98.3461 & - & - & - \\ 
        \hline
    \end{tabular}
    \label{tab:anova_cpu_usage}
\end{table*}

\subsubsection{Memory Usage Analysis}
\begin{figure}[!t]
    \centering
    \includegraphics[width=0.88\columnwidth]{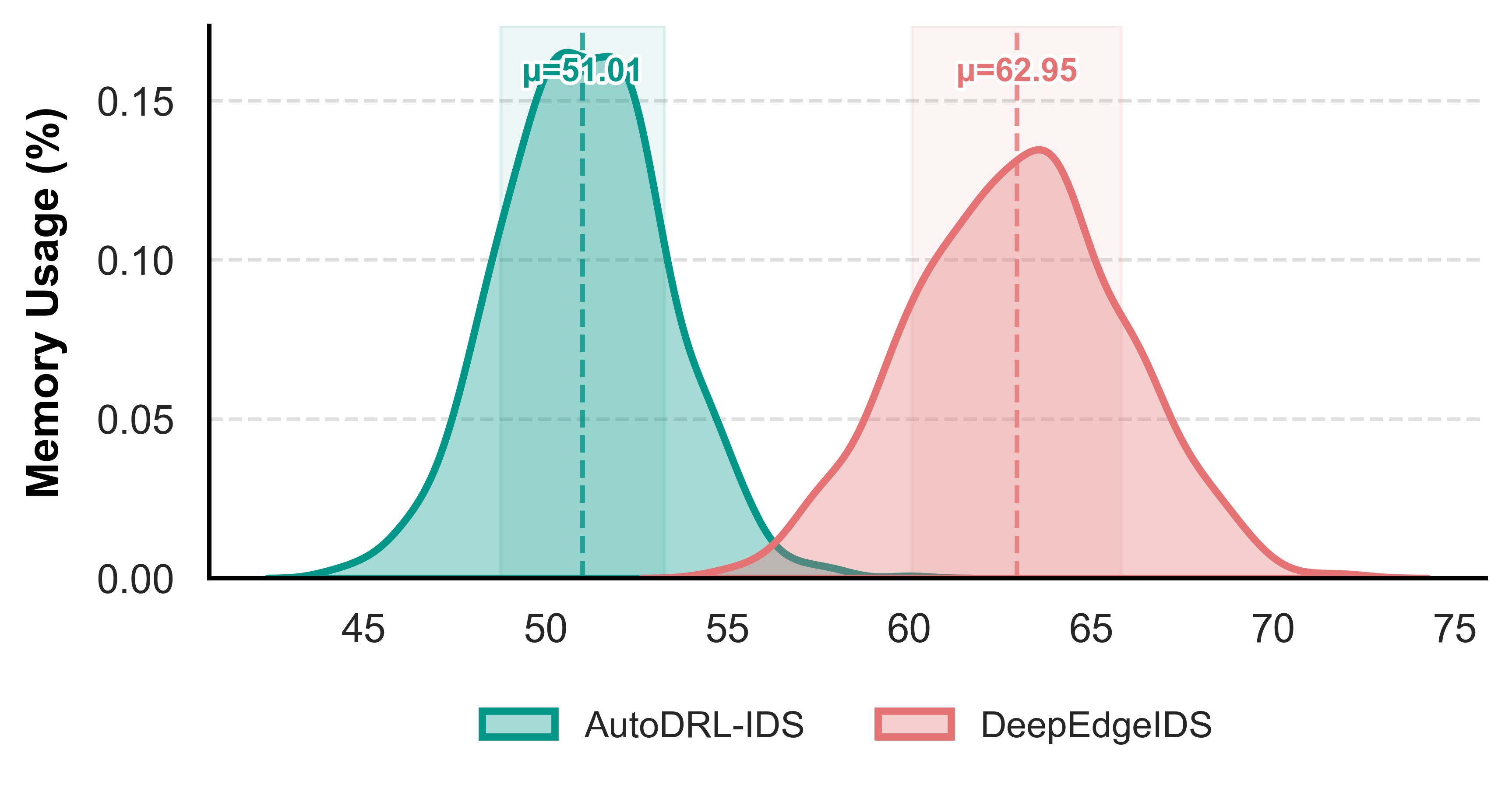}
    \caption{Memory usage comparison between AutoDRL-IDS and DeepEdgeIDS on edge gateways.}
    \label{fig:memory_usage}
\end{figure}

Figure~\ref{fig:memory_usage} compares the memory usage of AutoDRL-IDS and DeepEdgeIDS at the edge. DeepEdgeIDS exhibits a slightly higher median memory usage of around 60\%, while AutoDRL-IDS averages 50\%. This small difference reflects the adaptive behavior of DeepEdgeIDS, which continuously extracts features, interacts with the replay buffer, and refines policies to maintain real-time responsiveness. AutoDRL-IDS, by contrast, operates with a more static memory profile due to its reliance on pre-trained policies, resulting in greater consistency under steady workloads.\\
The ANOVA results in Table~\ref{tab:anova_memory_usage} indicate that the difference in memory usage between the two DRL-based IDS is not statistically significant (F = 2.18, P $>$ 0.05). This demonstrates that both DRL-based IDSs maintain comparable, efficient memory management suitable for edge gateways. Although DeepEdgeIDS requires more memory to support its self-adaptive learning operations, this increase is insignificant in both statistical and practical terms. Similar to the CPU trend, the slightly higher memory footprint of DeepEdgeIDS supports its adaptability and rapid per-decision response, without imposing a significant deployment burden on the edge gateway. Furthermore, both DeepEdgeIDS and AutoDRL-IDS maintain lightweight memory footprints, enabling real-time edge operations.
\begin{table*}[h]
    \centering
    \small
    \caption{ANOVA: Memory Usage Comparison Between DeepEdgeIDS and AutoDRL-IDS on Edge Gateway.}
    \begin{tabular}{|l|c|c|c|c|c|}
        \hline
        \textbf{Source} & \textbf{Degrees of Freedom} & \textbf{Sum of Squares} & \textbf{Mean Square} & \textbf{F Statistic} & \textbf{P-value} \\ 
        \hline
        Between Groups & 1 & 0.0124 & 0.0124 & 2.18 & $>$0.05 \\ 
        Within Groups & 38 & 0.2163 & 0.0057 & - & - \\ 
        Total & 39 & 0.2287 & - & - & - \\ 
        \hline
    \end{tabular}
    \label{tab:anova_memory_usage}
\end{table*}

\section{Discussion}
\label{Discussion}
The evaluation demonstrates that DeepEdgeIDS and AutoDRL-IDS exhibit distinct mathematical and operational behaviors, governed by their underlying learning paradigms. Their differences can be formally expressed in terms of optimization dynamics, convergence stability, and resource–performance trade-offs within the edge.
DeepEdgeIDS employs an unsupervised RL policy $\pi_{\theta}(a|s)$ that jointly minimizes a reconstruction error $L_r = \|x - \hat{x}\|_2^2$ and maximizes an expected reward $J(\pi_{\theta}) = \mathbb{E}_{s,a}[R_t]$. The gradient update follows $\nabla_{\theta} J(\pi_{\theta}) = \mathbb{E}[R_t \nabla_{\theta} \log \pi_{\theta}(a|s)]$, yielding a continuous feedback adaptation loop that maintains $\mathbb{E}[\|\nabla_{\theta} J\|^2] < \delta$, ensuring bounded learning variance. This mechanism enables real-time policy adjustments to shifting traffic distributions, thereby minimizing the detection error $E_d = P(\text{FN}) + P(\text{FP})$ and maintaining consistent performance in zero-day attack scenarios.
Conversely, AutoDRL-IDS optimizes a fixed Q-value regression objective $L_Q = (Q(s,a) - Q^*(s,a))^2$, constraining its updates to supervised targets. The reduction of stochastic gradient variance, $\text{Var}[\nabla_{\theta} J(\pi_{\theta})] \approx 0$, leads to deterministic inference but limits adaptation when traffic states deviate from the training distribution. Thus, while computationally efficient, AutoDRL-IDS exhibits lower entropy $H(\pi_{\theta})$ and slower convergence to unseen attack patterns than DeepEdgeIDS.
Latency behavior further reflects these learning dynamics. For DeepEdgeIDS, direct policy correction $\Delta \theta_t \propto \nabla_{\theta_t} (R_t - \bar{R})$ accelerates convergence, producing shorter cumulative adaptation cycles $T_c$. Formally, $T_{c,\text{DeepEdge}} < T_{c,\text{AutoDRL}}$, resulting in lower response time $\Delta t_r = t_{\text{AutoDRL}} - t_{\text{DeepEdge}} > 0$. This reduced temporal lag is consistent with its lower inference depth and adaptive local gradient propagation, which explain the observed mitigation speed during DDoS escalation phases.
Energy and computational efficiency metrics quantify the cost of this adaptivity. The expected energy consumption $E[P]$ scales with the update frequency $f_u$ and gradient magnitude $\|\nabla_{\theta} J(\pi_{\theta})\|$, such that $E[P] \propto f_u \cdot \mathbb{E}[\|\nabla_{\theta} J\|]$. Consequently, DeepEdgeIDS exhibits a higher mean power draw and a higher carbon footprint, $C_{\text{CO}_2}$, than AutoDRL-IDS. Nevertheless, this overhead directly correlates with enhanced robustness, reducing the successful attack ratio $\Gamma = A_{\text{success}} / A_{\text{total}}$. The energy–robustness correlation thus delineates a quantifiable exchange: higher $\nabla_{\theta}$ activity yields improved defensive plasticity at proportional energy cost. This relationship extends to a multi-objective optimization surface balancing detection accuracy $\alpha$, adaptability $\beta$, and sustainability $\sigma$:
\[
\max_{\pi_{\theta}} \{\alpha(\pi_{\theta}), \beta(\pi_{\theta})\} \quad \text{s.t.} \quad \sigma(\pi_{\theta}) \leq \sigma_{\max}.
\]
DeepEdgeIDS occupies a higher $(\alpha, \beta)$ region with an elevated $\sigma$, indicating adaptability at a moderate resource cost. AutoDRL-IDS lies closer to the efficiency-optimal frontier, maintaining a low $\sigma$ while reducing $\beta$ under dynamic loads. 
Additionally, the statistical outcomes, as evidenced by ANOVA significance and effect size analyses, confirm that DeepEdgeIDS achieves adaptability, faster response times, and higher detection accuracy through continuous gradient-driven learning, while AutoDRL-IDS provides deterministic stability and energy efficiency.
\section{Limitations and Future Work}
\label{Limitations and Future Work}
Despite its adaptability and resilience, DeepEdgeIDS faces several technical limitations that must be addressed for large-scale deployment. Its continuous learning and feature-extraction processes impose additional computational and energy overhead on the edge gateway, suggesting the need for optimization through asynchronous inference, model pruning, and adaptive update control. The system also remains partially vulnerable to adversarial perturbations that can destabilize its policy during dynamic traffic fluctuations; incorporating robust training and defensive regularization would mitigate this issue. 
Scalability poses another challenge, as coordinating distributed learning across multiple gateways demands efficient synchronization and bandwidth management. Federated and decentralized RL could enhance policy sharing while maintaining privacy and stability. In addition, DeepEdgeIDS’s detection precision comes at the cost of higher energy consumption, underscoring the need for future research into adaptive energy–performance balancing to ensure sustainable, context-aware operation in real-time IoT networks.

\section{Conclusion}
\label{Conclusion}
This study presents two DRL–based IDSs, DeepEdgeIDS and AutoDRL-IDS, which were evaluated within realistic edge networks under DDoS attacks. Results confirm that DeepEdgeIDS achieves adaptability, faster response times, and higher detection accuracy against evolving and zero-day attacks through unsupervised policy optimization. Its ability to autonomously refine decision boundaries enables resilient defense under dynamic network conditions. In contrast, AutoDRL-IDS attains competitive detection performance with lower computational and energy overhead, demonstrating strong suitability for constrained edge devices. The comparative analysis highlights a core trade-off between adaptability and efficiency: DeepEdgeIDS favors dynamic responsiveness, whereas AutoDRL-IDS emphasizes stability and sustainability.
Moreover, the findings establish DRL as a practical foundation for smart, edge-resident IDS. By achieving a balance between accuracy, resource efficiency, and robustness, these architectures advance the development of sustainable, self-adaptive security frameworks for next-generation IoT ecosystems.

\bibliographystyle{IEEEtran}
\bibliography{sample-base}

\end{document}